\documentclass[reqno]{amsart}
\usepackage{verbatim}

\usepackage{amsmath,amssymb,amsthm,cite,epsfig,graphics}
\usepackage{cite}

\newcommand{\p}{\partial}

\newcommand{\R}{{\mathbb R}}
\newcommand{\N}{{\mathbb N}}
\newcommand{\Z}{{\mathbb Z}}
\newcommand{\cw}{{\check\omega}}
\newcommand{\w}{{\omega}}
\def\r{\rangle}
\def\l{\langle}

\newtheorem*{lemma*}{Lemma}

\newtheorem*{conjecture*}{Conjecture}

{\theoremstyle{definition}

\newtheorem{example}{Example}
\newtheorem{remark}{Remark}

\newtheorem*{note*}{Note}
}

\allowdisplaybreaks

\begin{document}
\allowdisplaybreaks

\title[3D transforms]
{Three dimensional $C$-, $S$- and $E$-transforms}

\author{Maryna Nesterenko$^{1,2}$}
\author{Ji\v{r}\'{i} Patera$^1$}

\date{\today}

\begin{abstract}\
Three dimensional continuous and discrete Fourier-like transforms, based on the three simple and four semisimple compact Lie groups of rank 3, are presented. For each simple Lie group, there are three families of special functions ($C$-, $S$-, and $E$-functions) on which the transforms are built. Pertinent properties of the functions are described in detail, such as their orthogonality within each family, when integrated over a finite region $F$ of the 3-dimensional Euclidean space (continuous orthogonality), as well as when summed up over a lattice grid $F_M\subset F$ (discrete orthogonality).
The positive integer $M$ sets up the density of the lattice containing $F_M$. The expansion of functions given either on $F$ or on $F_M$ is the paper's main focus.
\end{abstract}

\maketitle

\noindent
$^1$ Centre de recherches math\'ematiques,
         Universit\'e de Montr\'eal,
         C.P.6128-Centre ville,
         Montr\'eal, H3C\,3J7, Qu\'ebec, Canada; patera@crm.umontreal.ca\\
$^2$ Institute of mathematics of NAS of Ukraine,
         3, Tereshchenkivs'ka str,
         Kyiv-4, 04216, Ukraine;\linebreak
         maryna@imath.kiev.ua

\noindent
\section{Introduction}\label{sec_Introduction}

New $n$-dimensional $C$- and $S$- and $E$-transforms were recently described in~\cite{KlimykPatera2006,KlimykPatera2007-1,KP3}. Each transform is based on a compact semisimple Lie group of rank $n$ and comes in three versions: analogs of Fourier series, Fourier integrals, and Fourier transforms on an $n$-dimensional lattice. They are named $C$- and $S$- and $E$-transforms \cite{patera2005} in recognition of the fact that they can be understood as generalizations of the one dimensional cosine, sine, and exponential Fourier transform.

The aim of this paper is to set the grounds for the 3-dimensional exploitation of transforms -- described here as continuous transforms in a finite region $F$ of the 3-dimensional Euclidean space $\R^3$, and also as discrete transforms of functions given on a lattice grid of points $F_M\subset F$ of any density -- in ready-to-use form. A positive integer $M$ specifies the density. In some cases, the density of the grid is more flexible as it is dictated by not one, but two or three positive intergers. The grid could thus be made denser along certain axes. The symmetry of the lattice is dictated by the shape of $F$, or equivalently, by the choice of the Lie group.

There are seven compact semisimple Lie groups of rank 3:
\begin{gather*}
SU(2)\times SU(2)\times SU(2)\,,\ \
SU(3)\times SU(2)\,,\ \
O(5)\times SU(2)\,,\ \
G(2)\times SU(2)\,,\\
SU(3)\,,\ \
O(7)\,,\ \
Sp(6)\,.
\end{gather*}
Throughout the paper we identify these cases by symbols that are often used for their respective Lie algebras:
$$
A_1\times A_1\times A_1\,,\quad
A_2\times A_1\,,\quad
C_2\times A_1\,,\quad
G_2\times A_1\,,\quad
A_3\,,\quad
B_3\,,\quad
C_3\,.
$$

The immediate motivation for this paper is our anticipation of the extensive use of the transforms given the need for processing the rapidly increasing amount of 3D digital data gathered today. In 2D, our group transforms offered only in some cases more than a marginal advantage,
having emerged when satisfactory practical methods had already been developed and adequately implemented. So far, practical use of the functions in 2D rested on the fact that the continuous extension of the transformed lattice data displayed remarkably smooth interpolation between lattice points \cite{GermainPatera2006-1} (see also references therein).

Special functions, which serve as the kernel of our transform (we call them $C$-, $S$-, and $E$-functions or orbit functions), have simple symmetry property under the action of the corresponding affine Weyl group. The affine group contains as a subgroup the group of translations in $\R^n$, which underlies the common Fourier transform. This is the primary reason for the superior performance of our transforms, although detailed comparisons, rather than examples, will have to provide quantitative content to substantiate such a claim.

Other properties of the $C$-, $S$-, and $E$-functions are not less important.

Within each family, functions are described  in a uniform way for semisimple Lie groups of any type and rank. In this work, we illustrate this uniformity by considering all seven rank 3 group cases in parallel. The price to pay for the uniformity of methods is having to work with non-orthogonal bases which are not normalized.

The functions are defined in $\R^n$ and have continuous derivatives of all degrees.
Their orthogonality, when integrated over the finite region $F$
appropriate for each Lie group, was shown in \cite{MoodyPatera2006}.
The discrete orthogonality of $C$-fun\-ctions in $F_M$
has already been described in \cite{MoodyPatera1987} and extensively used (see for example \cite{GrimmPatera1997}
and references therein).
The completeness of these systems of functions directly follows from the completeness of the system
of exponential functions.

A Laplace operator for each Lie group is given in a different set of coordinates.
The $C$- and $S$-functions are its eigenfunctions with known eigenvalues.
On the boundary of $F$, the $C$-functions have a vanishing normal derivative,
while $S$-functions reach zero at the boundary.

The functions have a number of other useful properties,
which can be found in~\cite{KlimykPatera2006,KlimykPatera2007-1,KP3}.
For example, the decomposition of their products into sums,
the splitting of functions into as many mutually exclusive congruence classes
as is the order of the center of the Lie group, etc$\dots$.

A different but valid viewpoint on some of the special functions presented here, namely, functions symmetrized by the summation of constituent functions over a finite group~\cite{Macdonald1995}, may turn out to be rather useful. The finite group, in the case of $C$- and $S$-functions, is the Weyl group of the corresponding semisimple Lie group. In the case of $E$-functions, it is the even subgroup of the Weyl group. The Weyl group of $SU(n)$ is isomorphic to the group $S_n$ of the permutation of $n$ elements. This led to the recent implementations in \cite{KlimykPatera2007-2, KlimykPatera2007-3}, where instead of the Weyl group of $SU(n)$, the $S_n$ group is used, and variables are given relative to an orthonormal system of coordinates. Furthermore, the even subgroup of $S_n$ is the alternating group. Related transforms were introduced most recently in~\cite{KP2008,KP6}.

The paper is organized as follows.

In Section~\ref{sec_properties_definitions}, necessary definitions and properties of Lie groups and algebras are given and discussed. Semisimple Lie groups of rank 3 are considered in detail in Section~\ref{sec_algebras}.
For each of these groups, we lay down the information
necessary to construct and use their orbit functions for 3D continuous and discrete transforms.
Section~\ref{sec_orbit-funcs} is devoted to $C$-, $S$- and $E$- orbit functions and to their pertinent properties. Continuous and discrete orbit-function transforms are presented in Section~\ref{sec_c-s-e-transforms}. Some problems and possible applications arising in connection with orbit functions are formulated in the conclusion. An example of the application of orbit function transforms in the case of the group $SU(2)\times SU(2)\times SU(2)$ is given at the end of the paper.

\section{Pertinent properties of Lie groups and Lie algebras}\label{sec_properties_definitions}
The notion of orbit function of $n$ variables depends essentially on the underlying semisimple Lie group of rank $n$. This section is intended to recall some of the standard properties of semisimple Lie groups/Lie algebras in general, and particularly those of rank 3, as well as properties of related Weyl groups. We also fix notation and terminology. Additional information about such Lie groups can be found for example in~\cite{Bourbaki, Humphreys1972,KassMoodyPateraSlansky1990, VinbergOnishchik}.

\subsection{Definitions and notations}\label{ssec_definitions}\

Let $\R^n$ be the real Euclidean space spanned by the simple roots of a~simple Lie group~$G$ (equivalently, Lie algebra). The basis of the simple roots is hereafter referred to as the $\alpha$-basis. An $\alpha$-basis is not orthogonal and comprises simple roots of at most two different lengths. If~a~semisimple $G$ is not simple, the $\alpha$-bases of its simple constituents are pairwise orthogonal.

For important practical (i.e. computational) reasons, it is advantageous to introduce also the basis of fundamental weights, hereafter referred to as the $\omega$-basis. Moreover, for Lie groups with simple roots of two different lengths, it is useful to introduce bases dual to $\alpha$- and $\omega$-bases, denoted here as $\check{\alpha}$- and $\check{\omega}$-bases respectively. Occasionally it is also useful to work with the orthonormal basis $\{e_1,e_2,\ldots e_n\}$ of $\R^n$. Each subsection contains an explicit elaboration of these bases for the groups we consider.

The Cartan matrix $C$ of $G$ provides, in principle, all of the information needed about $G$. It is an $n\times n$-matrix. In particular, it provides the relation between $\alpha$-and $\omega$-bases:
\begin{gather*}\label{bases}
\alpha=C\omega\qquad\qquad\omega=C^{-1}\alpha\,,
\end{gather*}
where
\begin{gather*}\label{Cartan_matrix}
C_{ij}=\left(
\tfrac{2\langle\alpha_i,\alpha_j\rangle}{\langle\alpha_j,\alpha_j\rangle}
\right),
\qquad i,j\in\{1, 2, \ldots, n\}.
\end{gather*}
Here $\langle\cdot\,,\cdot\rangle$ denotes the inner product in $\R^n$. The length of the long simple roots is determined by an additional convention
\begin{gather*}
\langle\alpha_{\text{long}} ,\alpha_{\text{long}}\rangle=2.
\end{gather*}

The dual bases are fixed by the relations
\begin{gather*}\label{duals}
\check{\alpha}_i=\frac{2\alpha_i}{\langle\alpha_i ,\alpha_i\rangle},
\qquad
\check{\omega}_i=\frac{2\omega_i}{\langle\alpha_i ,\alpha_i\rangle},
\qquad
\langle\alpha_i ,\check{\omega}_j\rangle=\langle\check{\alpha}_i ,\omega_j\rangle=\delta_{ij},
\end{gather*}
where $\delta_{ij}$ is the Kronecker delta.

The root lattice $Q$ and the weight lattice $P$  of $G$ are formed by all integer linear combinations of the $\alpha$-basis and $\omega$-basis,
\begin{gather*}
Q=\Z\alpha_1+\Z\alpha_2+\cdots+\Z\alpha_n,\qquad P=\Z\omega_1+\Z\omega_2+\cdots+\Z\omega_n.
\end{gather*}
In general, $Q\subseteq P$, but in rank 3 Lie groups the equalities do not occur. Similarly, we can introduce the dual lattices $\check Q$ and $\check P$.

In the weight lattice $P$, we define the cone of dominant weights $P^+$ and its subset of strictly dominant weights $P^{++}$
\begin{gather*}
P\;\supset\;
P^+=\Z^{\ge 0}\omega_1+\cdots+\Z^{\ge 0}\omega_n
\;\supset\;
P^{++}=\Z^{>0}\omega_1+\cdots+\Z^{>0}\omega_n.
\end{gather*}

For any simple Lie group $G$ there is a unique highest root $\xi$
\begin{gather}\label{highest_root}
\xi=m_1\alpha_1+m_2\alpha_2+\cdots+m_n\alpha_n=q_1\check\alpha_1+q_2\check\alpha_2+\cdots+q_n\check\alpha_n
\end{gather}
Coefficients $m_i$ and $q_i$, $i=\overline{1,n}$, are natural numbers, referred to as marks and comarks respectively. They are well known for all simple Lie groups (see for example \cite{Humphreys1972, KassMoodyPateraSlansky1990}).

\subsection{Weyl groups and their orbits}\label{ssec_Weyl_group}\

The Weyl group $W(G)$ of a semisimple Lie group $G$ is the finite group generated in $\R^n$ by reflections in $n-1$ dimensional mirrors (hyperplanes) orthogonal to the simple roots of $G$ and containing the origin of $\R^n$. For a simple root $\alpha_i$, $i=\overline{1,n}$  the corresponding reflection~$r_{\alpha_i}$ is given by
\begin{gather}\label{reflection_alpha}
r_i x=r_{\alpha_i} x=x-\frac{2\l x ,\alpha_i\r}{\l\alpha_i ,\alpha_i\r}\alpha_i
          =x-\l x ,\check\alpha_i\r\alpha_i,\qquad x\in\R^n.
\end{gather}

There is a general method for building Weyl group orbits. In Section~\ref{sec_algebras}, we limit ourselves to recording the result of its application for all Lie groups of rank $n=3$.

It is assumed that we have fixed a semisimple Lie group of rank 3 and that we consider its weight lattice $P$. A $W$-orbit can be generated from any point $(a,b,c)\in\R^3$, but in this paper we are almost always interested in $W$-orbits of points in $P$. It is convenient to specify the orbit by its unique point $(a,b,c)=a\omega_1+b\omega_2+c\omega_3$ with positive integer coordinates $a>0$, $b>0$, $c>0$. We denote such a generic orbit by $W(a, b, c)$.

Every group contains the trivial one point orbit $(0, 0, 0)$. The number of points of a generic orbit $|W(a, b, c)|$ is equal to the order $|W|$ of the Weyl group.

\subsection{Affine Weyl groups and their fundamental domains}\label{ssec_affine_weyl_group}\

Consider the reflection $r_{\xi}$ with respect to the hyperplane containing the origin and orthogonal to the highest root~$\xi$, see~(\ref{highest_root})
\begin{gather*}\label{reflection_xi}
r_{\xi} x=x-\frac{2\l x ,\xi\r}{\l\xi ,\xi\r}\xi,\qquad x\in\R^n.
\end{gather*}

We extend the set of  $n$ reflections $r_{\alpha_i}$, given in~(\ref{reflection_alpha}), generating the Weyl group $W$, by one reflection $r_0$
\begin{gather*}\label{reflection_0}
r_{0} x=r_\xi x+\check\xi,\qquad
\text{where}\quad \check\xi=\frac{2\xi}{\l\xi ,\xi\r},\quad x\in\R^n.
\end{gather*}
The resulting group transformations of $\R^n$, generated by  $n+1$ reflections $r_0,\ r_1,\ \ldots,\ r_n$, is referred to as the affine Weyl group $W^{aff}$. The order of  $W^{aff}$ is infinite.

The fundamental region $F(G)\subset \R^n$ for any $W^{aff}(G)$ is the convex hull of the vertices
$\{0,\frac{\omega_1}{q_1},\ldots, \frac{\omega_n}{q_n}\}$,
where $q_i$, $i=\overline{1,n}$ are comarks from~(\ref{highest_root}), or equivalently,
\begin{gather}\label{fund_region_Fm}
F(G)=\{0,\frac{\check\omega_1}{m_1},\frac{\check\omega_2}{m_2},\ldots, \frac{\check\omega_n}{m_n}\},
\qquad \text{where}\quad m_i,\; i=\overline{1,n}\quad\text{are marks, see~(\ref{highest_root}).}
\end{gather}

Repeated reflections of $F(G)$ in its $n-1$-dimensional sides results in tiling the entire space $\R^n$ by copies of $F$. We define grid $F_M\subset F$, depending on an arbitrary natural number $M$ as given in~(\ref{highest_root}),
\begin{gather*}\label{grid_Fm}
F_M=\left\{
\frac{s_1}{M}\check\omega_1+\frac{s_2}{M}\check\omega_2+\cdots+\frac{s_n}{M}\check\omega_n\mid
s_1,\dots,s_n\in\Z^{\geq0},\ \;
\sum_{i=1}^n s_im_i\le M>0
\right\}.
\end{gather*}

The number of points of the grid $F_M$ is denoted by $|F_M|$, and the volume of the fundamental region $F$ with respect to the euclidian measure is denoted by $|F|$.

\begin{remark}
In case $G=G_1\times G_2$, where $G_1$ and $G_2$ are simple, the fundamental region of $G$
is the Cartesian product of the fundamental regions of $G_1$ and $G_2$.
The same holds for grids $F_M$, but for each simple constituent of $G$,
the numbers $M_1$ nd $M_2$  could be chosen independently.

In the case of two different numbers $M_1$ and $M_2$, such that \mbox{$M_1\mod M_2=0$},
the corresponding grids are related as follows $F_{M_2}\subset
F_{M_1}$.
\end{remark}

\subsection{Even subgroup $W^e$ of the Weyl group}\label{ssec_even_subgroup}\

Elements of the subgroup $W^e\subset W$ are formed by an even number of reflections which generate $W$.
Each $C-$ and  $S-$function is built on a single Weyl group orbit of point $\lambda\in P$,
and each $E-$function is built on an orbit of the even subgroup of the Weyl group, defined as follows.

Consider a subgroup of W generated by an even number of reflections
\begin{gather}
W_e=\{r_{i_1}r_{i_2}\dots r_{i_{2k}}|k\in \N,\ r_{i_l}\in W,\ l=\overline{1,2k}\}.
\end{gather}
$W_e$ is a normal subgroup of the index 2 of the Weyl group, i.e. $2|W_e|=|W|$.
Let us denote the orbit of point $\lambda\in P$ with respect to the action of $W_e$ by $W_e(\lambda)$,
and the size of this orbit by $|W_e(\lambda)|$, then
\begin{gather}\label{even_subgr_of_W}
W(\lambda)=\left\{
\begin{array}{ll}
W_e(\lambda)\cup W_e(r_i\lambda),\; \text{for some}\;  r_i\in W,& \text{when}\; \lambda\in P^{++}\\
W_e(\lambda),& \text{when}\; \lambda\in P^{+}\setminus P^{++}
\end{array}
\right.
\end{gather}
and one of the following relations holds true
$|W_e(\lambda)|=\frac 12 |W(\lambda)|$, when $\lambda\in P^{++}$
or $|W_e(\lambda)|=|W(\lambda)|$, when $\lambda\in P^{+}\setminus P^{++}$.
Note that each orbit of the even subgroup $W_e$ contains exactly one point from
$P_e:=P^+\cup r_iP^{++}$, $r_i\in W$.

Similarly, we define the even affine Weyl subgroup
\begin{gather*}
W^{aff}_e=\{r_{i_1}r_{i_2}\dots r_{i_{2k}}|k\in \N,\ r_{i_l}\in W^{aff},\ l=\overline{1,2k}\},
\end{gather*}
and its fundamental region
\begin{gather}\label{fund_region_Fem}
F_e=F\cup r_iF,
\end{gather}
where $F$ is a fundamental region of $W^{aff}$~(\ref{fund_region_Fm}) and $r_i\in r_0,r_1,\dots,r_n$.

The same formula holds true for the grid on the fundamental region $F_e$
\begin{gather*}\label{grid_Fem}
F_{e\,M}=F_M\cup r_iF_M,\qquad r_i\in W^{aff}.
\end{gather*}

\begin{remark}
As it follows from~(\ref{fund_region_Fem}), the fundamental region of $W_e$ is not unique,
hence it can be chosen in such a way as to be convenient for a given application.
\end{remark}

\begin{example}\label{ex_a1}
Consider the rank one compact simple Lie group $SU(2)$
(the corresponding Lie algebra is~$A_1$).

The Cartan matrix in this case is the $1\times 1$ matrix $C=(2)$.
The root system consists of two roots $\pm\alpha$.
The root and weight lattices are formed by integer multiples
of the simple root and integer multiples of the fundamental weight $\omega$.
\begin{gather*}
Q=\{\Z\alpha\},\qquad P=\{\Z\omega\},
\qquad
\text{where}\quad \alpha=C\omega\quad \text{so that}\quad \omega=\tfrac12\alpha.
\end{gather*}
Therefore $P=Q\cup(Q+\omega)$.

The Weyl group $W$ of $A_1$ is of order 2.
It is a reflection group acting in $\R$. We have $W=\{ 1,-1\}$ and $W_e=\{ 1\}$.
Consequently, a Weyl group orbit containing the point~$x\ne 0$, also contains the point~$-x$,
but the orbit of the even subgroup of $W$ consists of a single point, either $x$ or $-x$.

The fundamental region $F$ is the segment with endpoints $F=\{0,\omega\}$.

The grid $F_M\subset F$ is fixed by the positive integer $M$ and it consists of $M+1$ points
\begin{gather*}
F_M=\{0,\tfrac1M,\tfrac2M,\dots,\tfrac{M-1}{M},1\}.
\end{gather*}
\end{example}

\section{Semisimple Lie algebras of rank three}\label{sec_algebras}

In this section, we provide specific information about compact semisimple Lie groups of rank 3,
namely $SU(2)\times SU(2)\times SU(2)$, $O(5)\times SU(2)$, $SU(3)\times SU(2)$, $G_2\times SU(2)$,
and the simple groups $Sp(6)$, $O(7)$, and $SU(4)$.
For each of the groups ,we lay down all of the information necessary to construct and use their orbit functions for 3D continuous and discrete transforms.

Note, that only the weight lattices of the Lie groups
$SU(2)\times SU(2)\times SU(2)$, $O(5)\times SU(2)$, $O(7)$ and $Sp(6)$
display cubic symmetries.

Below we present the Dynkin diagram for each semisimple Lie group of rank 3.
On these diagrams, long and short simple roots $\alpha$
are respectively denoted by unfilled and filled circles
and comarks are presented over the circles.



\bigskip

\subsection{The Lie algebra ${A_1\times A_1\times A_1}$}\label{ssec_A1A1A1}\

\begin{figure}[h]
\centerline{\includegraphics[scale=0.88]{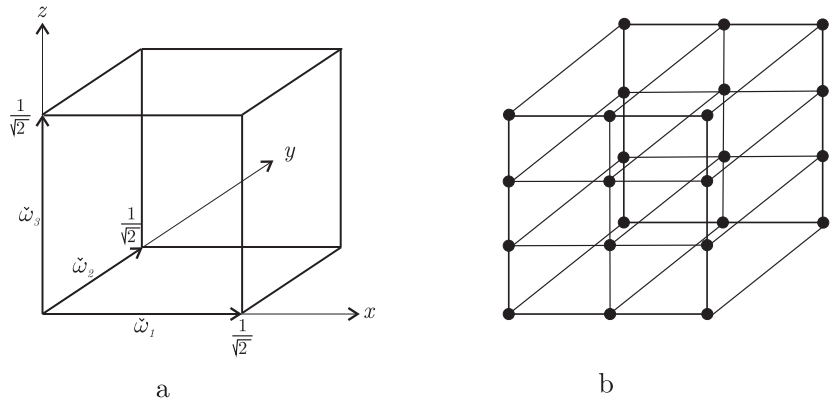}}
\caption{a)~the fundamental region $F$ of the Lie algebra $A_1\times A_1\times A_1$;
$x$, $y$ and $z$ indicate
respectively the orthogonal directions of the orthonormal basis $\{e_1$, $e_2$, $e_3\}$;
b) the grid $F_{2,1,3}(A_1\times A_1\times A_1)$.}\label{fig_a1a1a1}
\end{figure}

This case is a straightforward concatenation of three copies of $A_1$ (see Example~\ref{ex_a1}).
The Dynkin diagram and Cartan matrix with its inverse are the following,
\begin{gather*}
\parbox{.6\linewidth}
{\setlength{\unitlength}{2pt}
\def\kr{\circle{5}}
\def\cr{\circle*{5}}
\thicklines
\begin{picture}(10,15)
\put(4,12){$1$}
\put(5,7){\kr}
\put(3,0){$\alpha_1$}
\put(14,12){$1$}
\put(15,7){\kr}
\put(13,0){$\alpha_2$}
\put(24,12){$1$}
\put(25,7){\kr}
\put(23,0){$\alpha_3$}
\end{picture}}
\hspace{-175 pt}
C=\left(\begin{smallmatrix}
                          2&0&0\\0&2&0\\0&0&2
                     \end{smallmatrix}\right),
\qquad
C^{-1}=\tfrac12\left(\begin{smallmatrix}
                          1&0&0\\0&1&0\\0&0&1
                     \end{smallmatrix}\right).
\end{gather*}

All simple roots of the same length equal to~$\sqrt 2$.
The bases of simple roots and fundamental weights are thus related by
\begin{gather*}
\begin{array}{lllll}
\alpha_i=2\omega_i,&\quad
\omega_i=\tfrac12\alpha_i,&\quad
\check\alpha_1=\alpha_1,&\quad
\check\omega_1=\omega_1,&\quad
i=1,2,3.
\end{array}
\end{gather*}

These bases are easily written in the orthonormal basis $\{e_1,\ e_2,\ e_3\}$
\begin{gather*}
\begin{array}{llll}
\alpha_i=\sqrt 2 e_i,&\quad
\omega_i=\frac{1}{\sqrt 2}e_i,&\quad
\check\omega_i=\frac{1}{\sqrt 2}e_i,&\quad
i=1,2,3.
\end{array}
\end{gather*}

The highest roots for the simple subgroups $A_1$ are given by the formulas
\begin{gather*}
\xi=\alpha_1,
\quad
\xi=\alpha_2,
\quad
\xi=\alpha_3.
\end{gather*}

The fundamental region is a cube with vertices, see Fig.~\ref{fig_a1a1a1}a
\begin{gather*}
F(A_1\times A_1\times A_1)=\{0,\ \omega_i,\ \omega_i+\omega_j,\ \omega_1+\omega_2+\omega_3\},
\quad \text{where} \quad j<i\in{1,2,3}.
\end{gather*}

The volume of the fundamental region is given by the formula
\begin{gather*}
|F(A_1\times A_1\times A_1)|=|\cw_1|\cdot|\cw_2|\cdot|\cw_3|=\frac{1}{2\sqrt 2}.
\end{gather*}

The grid $F_{M,M',M''}\subset F$ is fixed by the independent choice of
three positive integers $M,M'$ and $M''$, and consists of all the points
\begin{gather*}
F_{M,M',M''}(A_1\times A_1\times A_1)=\left\{\tfrac{s_1}{M}\cw_1+\tfrac{s'_1}{M'}\cw_2+\tfrac{s''_1}{M''}\cw_3\mid\
s_1\leq M,\  s'_1\leq M',\  s''_1\leq M';\ s_1,s'_1,s''_1\in \Z^{\ge 0}\right\}.
\end{gather*}
The grid is cubic if $M=M'=M''$, otherwise it is rectangular. The
freedom to use unequal values of $M,M'$ and $M''$ may prove
rather useful in the analysis of data given on rectangular, but not
cubic grids.

The number of points in the grid $F_{M,M',M''}$ equals to
\begin{gather*}
|F_{M,M',M''}(A_1\times A_1\times A_1)|=|F_{M}(A_1)|\cdot|F_{M'}(A_1)|\cdot|F_{M''}(A_1)|=(M+1)(M'+1)(M''+1).
\end{gather*}

\begin{example}\label{ex_grid_a1a1a1}\

Consider the case $M=M'=1$, $M''=3$. There are 16 points of $F_{1,1,3}(A_1\times A_1\times A_1)$.
Explicitly, we have the following sets of integers $[s_1,s_1',s_1'']$
and the corresponding grid points in the $\cw$-basis (in this case the basis is orthogonal),
$(\tfrac{s_1}M,\tfrac{s_1'}{M'},\tfrac{s_1''}{M''})$:
\begin{gather*}
\begin{array}{llll}
[0,0,0]=(0,0,0),&\;
[0,0,1]=(0,0,\tfrac13),&\;
[0,0,2]=(0,0,\tfrac23),&\;
[0,0,3]=(0,0,1),
\\[1pt]
[0,1,0]=(0,1,0),&\;
[0,1,1]=(0,1,\tfrac13),&\;
[0,1,2]=(0,1,\tfrac23),&\;
[0,1,3]=(0,1,1),
\\[1pt]
[1,0,0]=(1,0,0),&\;
[1,0,1]=(1,0,\tfrac13),&\;
[1,0,2]=(1,0,\tfrac23),&\;
[1,0,3]=(1,0,1),
\\[1pt]
[1,1,0]=(1,1,0),&\;
[1,1,1]=(1,1,\tfrac13),&\;
[1,1,2]=(1,1,\tfrac23),&\;
[1,1,3]=(1,1,1).
\end{array}
\end{gather*}

Next, consider the case  $M=2$, $M'=1$, $M''=3$, see Fig.~\ref{fig_a1a1a1}b.
The 16 points  of $F_{1,1,3}(A_1\times A_1\times A_1)$ form a part of $F_{2,1,3}(A_1\times A_1\times A_1)$.
Note that they correspond to another set of $[s_1,s_1',s_1'']$.
More precisely, $F_{1,1,3}$ is the subset of $F_{2,1,3}$  with an even value of $s_1$.
Thus $F_{2,1,3}(A_1\times A_1\times A_1)$ contains an additional eight points:
\begin{gather*}
\begin{array}{llll}
[1,0,0]=(\tfrac12,0,0),&\;
[1,0,1]=(\tfrac12,0,\tfrac13),&\;
[1,0,2]=(\tfrac12,0,\tfrac23),&\;
[1,0,3]=(\tfrac12,0,1),
\\[1pt]
[1,1,0]=(\tfrac12,1,0),&\;
[1,1,1]=(\tfrac12,1,\tfrac13),&\;
[1,1,2]=(\tfrac12,1,\tfrac23),&\;
[1,1,3]=(\tfrac12,1,1).
\end{array}
\end{gather*}
\end{example}

The Weyl group orbit of the generic point $a\omega_1+b\omega_2+c\omega_3$, $\ a,b,c>0$,
always consists of the eight points
\begin{gather*}
W_{(a,b,c)}(A_1\times A_1\times A_1)=\{(\pm a,\ \pm b,\ \pm c)\}.
\end{gather*}

Orbit sizes for arbitrary points are given by the relations
\begin{gather*}
\begin{array}{ll}
|W_{(a,b,c)}|=8,&\quad
|W_{(a,b,0)}|=|W_{(a,0,c)}|=|W_{(0,b,c)}|=4,
\\[1pt]
|W_{(0,0,0)}|=1,&\quad
|W_{(a,0,0)}|=|W_{(0,b,0)}|=|W_{(0,0,c)}|=2.
\end{array}
\end{gather*}


\bigskip

\subsection{The Lie algebra ${A_2\times A_1}$}\label{ssec_A2A1}\

\begin{figure}[ht]
\centerline{\includegraphics[scale=0.88]{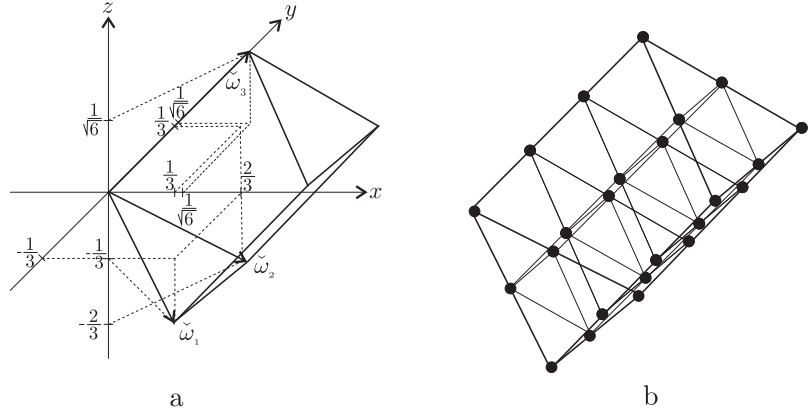}}
\caption{a)~the fundamental region $F$ of the Lie algebra $A_2\times A_1$;
$x$, $y$ and $z$ indicate respectively orthogonal directions of the orthonormal basis $\{e_1$, $e_2$, $e_3\}$
and the fundamental region of $A_2$ lies in the plane $x+y+z=0$;
b) the grid $F_{2,3}(A_2\times A_1)$.}\label{fig_a2a1}
\end{figure}

The Dynkin diagram and Cartan matrix with its inverse are the following,
\begin{gather*}
\parbox{.6\linewidth}
{\setlength{\unitlength}{1pt}
\def\kr{\circle{10}}
\def\cr{\circle*{10}}
\thicklines
\begin{picture}(20,30)
\put(8,24){$1$}
\put(10,14){\kr}
\put(6,0){$\alpha_1$}
\put(15,14){\line(1,0){10}}
\put(28,24){$1$}
\put(30,14){\kr}
\put(26,0){$\alpha_2$}
\put(48,24){$1$}
\put(50,14){\kr}
\put(46,0){$\alpha_3$}
\end{picture}}
\hspace{-175 pt}
C=\left(\begin{smallmatrix}
                          2&-1&0\\-1&2&0\\0&0&2
                     \end{smallmatrix}\right),
\qquad
C^{-1}=\tfrac16\left(\begin{smallmatrix}
                          4&2&0\\2&4&0\\0&0&3
                     \end{smallmatrix}\right).
\end{gather*}

Hence all the simple roots of the same length equal to~$\sqrt 2$.

The bases of simple roots and fundamental weights are thus related by
\begin{gather*}
\begin{array}{llll}
\alpha_1=2\omega_1-\omega_2,&\quad
\omega_1=\tfrac23\alpha_1+\tfrac13\alpha_2,&\quad
\check\alpha_1=\alpha_1,&\quad
\check\omega_1=\omega_1,
\\
\alpha_2=-\omega_1+2\omega_2,&\quad
\omega_2=\tfrac13\alpha_1+\tfrac23\alpha_2,&\quad
\check\alpha_2=\alpha_2,&\quad
\check\omega_2=\omega_2,
\\
\alpha_3=2\omega_3;&\quad
\omega_3=\tfrac12\alpha_3,&\quad
\check\alpha_3=\alpha_3,&\quad
\check\omega_3=\omega_3.
\end{array}
\end{gather*}

In order to visualize the implied geometry, it is useful to
represent the $\omega$ and $\alpha$ bases in the orthonormal basis.
We have
\begin{gather*}
\begin{array}{ll}
\alpha_1=(1,\ -1,\ 0)=e_1-e_2,&\quad
\omega_1=(\tfrac23,-\tfrac13,-\tfrac13)=\tfrac23e_1-\tfrac13e_2-\tfrac13e_3=\cw_1,
\\[1 pt]
\alpha_2=(0,\ 1,\ -1)=e_2-e_3,&\quad
\omega_2=(\tfrac13,-\tfrac13,-\tfrac23)=\tfrac13e_1+\tfrac13e_2-\tfrac23e_3=\cw_2,
\\[1 pt]
\alpha_3=\tfrac{\sqrt{2}}{\sqrt{3}}(1,1,1)=\tfrac{\sqrt{2}}{\sqrt{3}}(e_1+e_2+e_3),&\quad
\omega_3=\tfrac{1}{\sqrt6}(1,\ 1,\ 1)=\tfrac{1}{\sqrt6}(e_1+e_2+e_3)=\cw_3.
\end{array}
\end{gather*}

The highest roots for the simple subgroups $A_2$ and $A_1$ are given by the formulas
\begin{gather*}
\xi=\alpha_1+\alpha_2,
\qquad
\xi=\alpha_3.
\end{gather*}

The fundamental region $F$ is a cylinder (see Fig.~\ref{fig_a1a1a1}a)
with an equilateral triangle as its base and $\w_3$ as its height.
Its~vertices~are
\begin{gather*}
F(A_2\times A_1)=\{0,\ \w_1,\ \w_2,\ \w_3,\ \w_1+\w_3,\ \w_2+\w_3\}.
\end{gather*}

The volume of the fundamental region is given by the formula
\begin{gather*}
|F(A_2\times A_1)|=\tfrac12([\cw_1,\cw_2],\cw_3)=\tfrac12\cdot\tfrac13\cdot\tfrac13\cdot\tfrac{1}{\sqrt{6}}
\left|\begin{smallmatrix}
2&-1&-1\\1&1&-2\\1&1&1\\
\end{smallmatrix}\right|=\tfrac{1}{2\sqrt{6}}
\end{gather*}

The grid $F_{M,M'}(A_2\times A_1)\subset F(A_2\times A_1)$ is fixed by the independent choice of the
two positive integers $M$ and $M'$.
It consists of all the points
\begin{gather*}
F_{M,M'}(A_2\times A_1)=\left\{\tfrac{s_1}{M}\cw_1+\tfrac{s_2}{M}\cw_2+\tfrac{s'_1}{M'}\cw_3\mid\
s_1+s_2 \le M,\ s'_1 \le M';\ s_1,s_2,s_1'\in \Z^{\ge 0}\right\}.
\end{gather*}

The number of points in the grid $F_{M,M'}(A_2\times A_1)$ equals to
\begin{gather*}
|F_{M,M'}(A_2\times A_1)|=|F_{M}(A_2)|\cdot|F_{M'}(A_1)|=\left((M+1)^2-\tfrac {M(M+1)}{2}\right)(M'+1).
\end{gather*}

\begin{example}\label{ex_grid_a2a1}
Consider the case $M=2$, $M'=3$, see Fig.~\ref{fig_a2a1}b.
There exist 24 points of $F_{2,3}$
\begin{gather*}
|F_{2,3}(A_2\times A_1)|=\left((2+1)^2-\tfrac {2(2+1)}{2}\right)(3+1)=24.
\end{gather*}

Explicitly, we have the following sets of integers $[s_1,s_2,s_1']$
and the corresponding grid points $(\tfrac{s_1}M,\tfrac{s_2}{M},\tfrac{s_1'}{M'})$ in $\cw$-basis:
\begin{gather*}
\begin{array}{llll}
[0,2,0]=(0,1,0),&\;
[0,2,3]=(0,1,1),&\;
[0,2,2]=(0,1,\tfrac23),&\;
[0,2,1]=(0,1,\tfrac13),
\\[1pt]
[2,0,0]=(1,0,0),&\;
[2,0,3]=(1,0,1),&\;
[2,0,2]=(1,0,\tfrac23),&\;
[2,0,1]=(1,0,\tfrac13),
\\[1pt]
[0,0,0]=(0,0,0),&\;
[0,0,3]=(0,0,1),&\;
[0,0,2]=(0,0,\tfrac23),&\;
[0,0,1]=(0,0,\tfrac13),
\\[1pt]
[0,1,0]=(0,\tfrac12,0),&\;
[0,1,3]=(0,\tfrac12,1),&\;
[0,1,2]=(0,\tfrac12,\tfrac23),&\;
[0,1,1]=(0,\tfrac12,\tfrac13),
\\[1pt]
[1,0,0]=(\tfrac12,0,0),&\;
[1,0,3]=(\tfrac12,0,1),&\;
[1,0,2]=(\tfrac12,0,\tfrac23),&\;
[1,0,1]=(\tfrac12,0,\tfrac13),
\\[1pt]
[1,1,0]=(\tfrac12,\tfrac12,0),&\;
[1,1,3]=(\tfrac12,\tfrac12,1),&\;
[1,1,2]=(\tfrac12,\tfrac12,\tfrac23),&\;
[1,1,1]=(\tfrac12,\tfrac12,\tfrac13).
\end{array}
\end{gather*}
\end{example}

The Weyl group orbit of the generic point $a\omega_1+b\omega_2+c\omega_3$, $\ a,b,c>0$,
consists of twelve points
\begin{gather*}
W_{(a,b,c)}((A_2\times A_1))=
\{
(a,\ b,\ \pm c),\
(-a,\ a+b,\ \pm c),\
(a+b,\ -b,\ \pm c),\
(b,\ -(a+b),\ \pm c),
\\
(-(a+b),\ a,\ \pm c),\
(-b,\ -a,\ \pm c)
\}.
\end{gather*}

Orbit sizes for arbitrary points are given by the relations
\begin{gather*}
\begin{array}{llll}
|W_{(a,b,c)}|=12,&\quad
|W_{(a,b,0)}|=6,&\quad
|W_{(a,0,c)}|=6,&\quad
|W_{(0,b,c)}|=6,
\\[1pt]
|W_{(a,0,0)}|=3,&\quad
|W_{(0,b,0)}|=3,&\quad
|W_{(0,0,c)}|=2,&\quad
|W_{(0,0,0)}|=1.
\end{array}
\end{gather*}


\bigskip

\subsection{The Lie algebra ${C_2\times A_1}$}\label{ssec_C2A1}\

\begin{figure}[h]
\centerline{\includegraphics[scale=1]{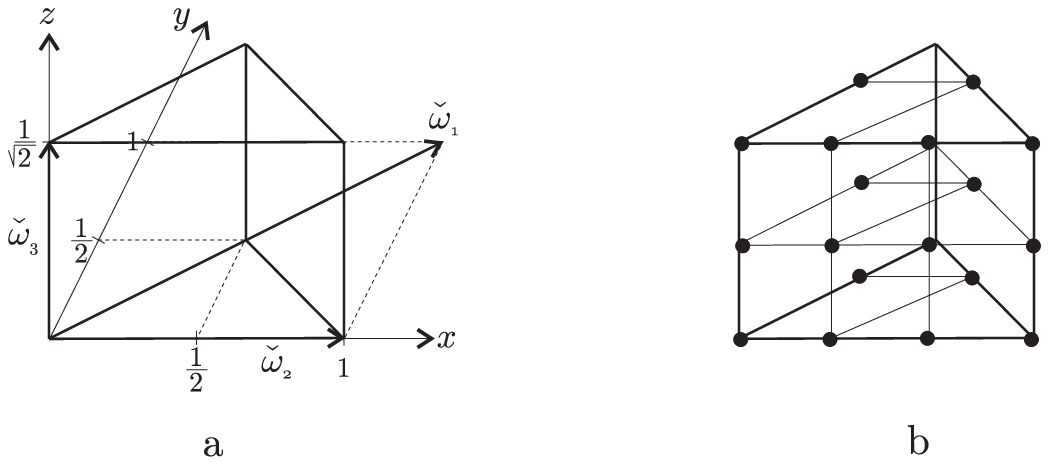}}
\caption{a)~the fundamental region $F$ of the Lie algebra $C_2\times A_1$;
$x$, $y$ and $z$ indicate respectively orthogonal directions of the orthonormal basis $\{e_1$, $e_2$, $e_3\}$
and the fundamental region of $C_2$ is the isosceles right-angled triangle in the plane $z=0$,
its right-angle is at the point $\tfrac12\cw_1$;
b) the grid $F_{3,2}(C_2\times A_1)$.}\label{fig_c2a1}
\end{figure}

The Dynkin diagram and Cartan matrix with its inverse are the following,
\begin{gather*}
\parbox{.6\linewidth}
{\setlength{\unitlength}{2pt}
\def\kr{\circle{5}}
\def\cr{\circle*{5}}
\thicklines
\begin{picture}(10,15)
\put(4,12){$2$}
\put(5,7){\cr}
\put(3,0){$\alpha_1$}
\put(6,5){\line(1,0){7,5}}
\put(6,9){\line(1,0){7,5}}
\put(14,12){$1$}
\put(15,7){\kr}
\put(13,0){$\alpha_2$}
\put(24,12){$1$}
\put(25,7){\kr}
\put(23,0){$\alpha_3$}
\end{picture}}
\hspace{-175 pt}
C=\left(\begin{smallmatrix}
                          2&-1&0\\-2&2&0\\0&0&2
                     \end{smallmatrix}\right),
\qquad
C^{-1}=\tfrac12\left(\begin{smallmatrix}
                          2&1&0\\2&2&0\\0&0&1
                     \end{smallmatrix}\right).
\end{gather*}

Hence $\alpha_1$ is the shorter of the simple roots.
The relative lengths of the simple roots are set as $\langle\alpha_1 ,\alpha_1\rangle=1$ and
$\langle\alpha_2 ,\alpha_2\rangle=\langle\alpha_3 ,\alpha_3\rangle=2$.

The bases of simple roots and fundamental weights are thus related by
\begin{gather*}
\begin{array}{llll}
\alpha_1=2\omega_1-\omega_2,&\quad
\omega_1=\alpha_1+\tfrac12\alpha_2,&\quad
\check\alpha_1=2\alpha_1,&\quad
\check\omega_1=2\omega_1,
\\
\alpha_2=-2\omega_1+2\omega_2,&\quad
\omega_2=\alpha_1+\alpha_2,&\quad
\check\alpha_2=\alpha_2,&\quad
\check\omega_2=\omega_2,
\\
\alpha_3=2\omega_3;&\quad
\omega_3=\tfrac12\alpha_3;&\quad
\check\alpha_3=\alpha_3;&\quad
\check\omega_3=\omega_3.
\end{array}
\end{gather*}

In the orthonormal basis these bases have the form
\begin{gather*}
\begin{array}{llll}
\alpha_1=(0,1,0)=e_2,&\quad
\omega_1=(\tfrac 12,\tfrac 12,0)=\tfrac 12e_1+\tfrac 12e_2,&\quad
\check\omega_1=(1,1,1),
\\
\alpha_2=(1,-1,0)=e_1-e_2,&\quad
\omega_2=(1,0,0)=e_1,&\quad
\check\omega_2=(1,0,0),
\\
\alpha_3=(0,0,\sqrt 2)=\sqrt 2 e_3;&\quad
\omega_3=(0,0,\tfrac{1}{\sqrt 2})=\tfrac {1}{\sqrt 2}e_3;&\quad
\check\omega_3=(0,0,\tfrac {1}{\sqrt 2}).
\end{array}
\end{gather*}

The highest roots for the simple subgroups $C_2$ and $A_1$ are given by the formulas
\begin{gather*}
\xi=2\alpha_1+\alpha_2,
\qquad
\xi=\alpha_3.
\end{gather*}

The fundamental region $F(C_2\times A_1)$ is a cylinder (see Fig.~\ref{fig_c2a1}a) with a triangular base.
Its~vertices~are
\begin{gather*}
F(C_2\times A_1)=\{0,\ \tfrac12 \cw_1,\ \cw_2,\ \cw_3,\ \tfrac12\cw_1+\cw_3,\ \cw_2+\cw_3\}.
\end{gather*}

The volume of the fundamental region is given by the formula
\begin{gather*}
|F(C_2\times A_1)|=\tfrac14([\cw_2,\cw_1],\cw_3)=\tfrac14\cdot\tfrac{1}{\sqrt{2}}
\left|\begin{smallmatrix}
1&0&0\\
1&1&0\\
0&0&1\\
\end{smallmatrix}\right|=\tfrac{1}{4\sqrt{2}}
\end{gather*}

The grid $F_{M,M'}\subset F$ is fixed by the independent choice of the
two positive integers $M$ and $M'$.
It consists of all the points
\begin{gather*}
F_{M,M'}(C_2\times A_1)=\left\{\tfrac{s_1}{M}\cw_1+\tfrac{s_2}{M}\cw_2+\tfrac{s'_1}{M'}\cw_3\mid\
2s_1+s_2  \le M,\ s'_1 \le M';\ s_1,s_2,s_1'\in \Z^{\ge 0}\right\}.
\end{gather*}

The number of points in the grid $F_{M,M'}(C_2\times A_1)$ equals to
\begin{gather*}
|F_{M,M'}(C_2\times A_1)|=|F_{M}(C_2)|\cdot|F_{M'}(A_1)|=
\left(\left[\tfrac{M}{2}\right]+1\right)
\left(M+1-\left[\tfrac{M}{2}\right]\right)
(M'+1),
\end{gather*}
where $[\ \cdot\ ]$ denotes the integer part of a number.

\begin{example}\label{ex_grid_c2a1}
Consider the case $M=3$, $M'=2$. There exist 18 points of $F_{3,2}(C_2\times A_1)$, see Fig.~\ref{fig_c2a1}b
\begin{gather*}
|F_{3,2}(C_2\times A_1)|=\left(\left[\tfrac{3}{2}\right]+1\right)\left(3+1-\left[\tfrac{3}{2}\right]\right)(2+1)=18.
\end{gather*}

Explicitly, we have the following sets of integers $[s_1,s_2,s_1']$
and the corresponding grid points in the $\cw$-basis,
$(\tfrac{s_1}M,\tfrac{s_2}{M},\tfrac{s_1'}{M'})$:
\begin{gather*}
\begin{array}{lll}
[1,0,0]=(\tfrac13,0,0),&\quad
[1,0,1]=(\tfrac13,0,\tfrac12),&\quad
[1,0,2]=(\tfrac13,0,2),
\\[1pt]
[0,1,0]=(0,\tfrac13,0),&\quad
[0,1,1]=(0,\tfrac13,\tfrac12),&\quad
[0,1,2]=(0,\tfrac13,2),
\\[1pt]
[0,2,0]=(0,\tfrac23,0),&\quad
[0,2,1]=(0,\tfrac23,\tfrac12),&\quad
[0,2,2]=(0,\tfrac23,2),
\\[1pt]
[1,1,0]=(\tfrac13,\tfrac13,0),&\quad
[1,1,1]=(\tfrac13,\tfrac13,\tfrac12),&\quad
[1,1,2]=(\tfrac13,\tfrac13,2),
\\[1pt]
[0,0,0]=(0,0,0),&\quad
[0,0,1]=(0,0,\tfrac12),&\quad
[0,0,2]=(0,0,2),
\\[1pt]
[0,3,0]=(0,1,0),&\quad
[0,3,1]=(0,1,\tfrac12),&\quad
[0,3,2]=(0,1,2).
\end{array}
\end{gather*}
\end{example}

The Weyl group orbit of the generic point $a\omega_1+b\omega_2+c\omega_3$, $\ a,b,c>0$,
consists of the following set of points
\begin{gather*}
W_{(a,b,c)}(C_2\times A_1)=\{
\pm(a, b, c),\
\pm(a, b, -c),\
\pm(-a, a+b, c),\
\pm(a+2b, -b, c),\
\pm(a+2b, -(a+b), c),
\\ \qquad\;
\pm(-a, a+b, -c),\
\pm(a+2b, -(a+b), -c),\
\pm(a+2b, -b, -c)
\}.
\end{gather*}

Orbit sizes for arbitrary points are given by the relations
\begin{gather*}
\begin{array}{llll}
|W_{(a,b,c)}|=16,&\quad
|W_{(a,b,0)}|=8,&\quad
|W_{(a,0,c)}|=8,&\quad
|W_{(0,b,c)}|=8,
\\[1pt]
|W_{(a,0,0)}|=4,&\quad
|W_{(0,b,0)}|=4,&\quad
|W_{(0,0,c)}|=2,&\quad
|W_{(0,0,0)}|=1.
\end{array}
\end{gather*}


\bigskip

\subsection{The Lie algebra ${G_2\times A_1}$}\label{ssec_G2A1}\

\begin{figure}[h]
\centerline{\includegraphics[scale=0.88]{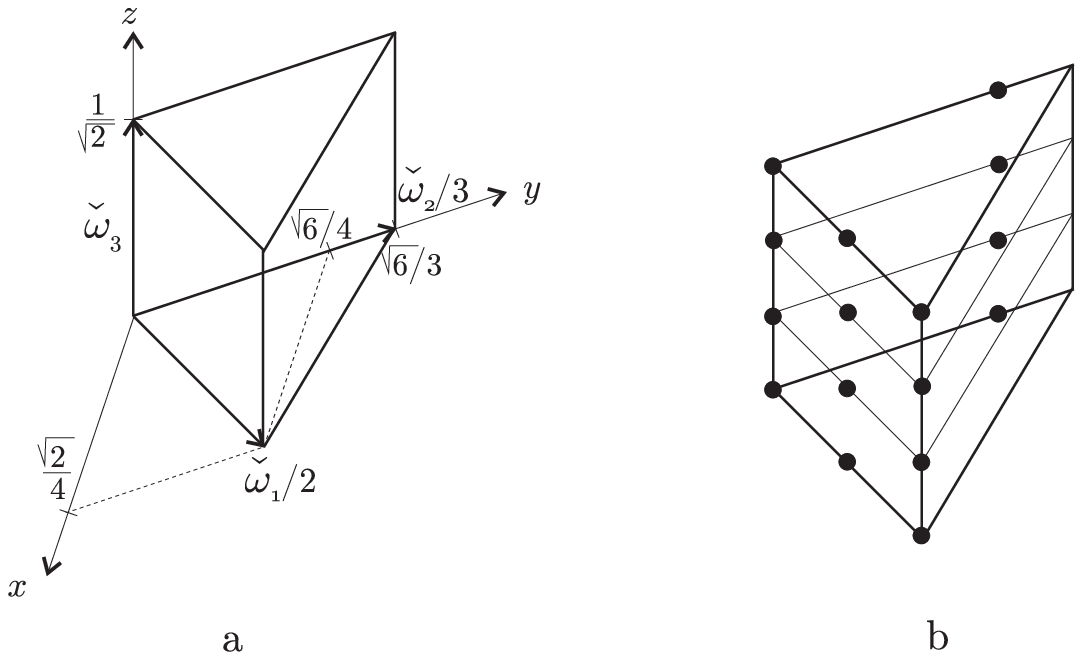}}
\caption{a)~the fundamental region $F$ of the Lie algebra $G_2\times A_1$;
$x$, $y$ and $z$ indicate respectively orthogonal directions of the orthonormal basis $\{e_1$, $e_2$, $e_3\}$
and the fundamental region of $G_2$ is the half of the equilateral triangle in the plane $z=0$;
b) the grid $F_{4,3}(G_2\times A_1)$.}\label{fig_g2a1}
\end{figure}

The Dynkin diagram and Cartan matrix with its inverse are the following,
\begin{gather*}
\parbox{.6\linewidth}
{\setlength{\unitlength}{1pt}
\def\kr{\circle{10}}
\def\cr{\circle*{10}}
\thicklines
\begin{picture}(20,30)
\put(8,24){$2$}
\put(10,14){\kr}
\put(6,0){$\alpha_1$}
\put(13,10){\line(1,0){18}}
\put(15,14){\line(1,0){12}}
\put(13,18){\line(1,0){18}}
\put(28,24){$3$}
\put(30,14){\cr}
\put(26,0){$\alpha_2$}
\put(48,24){$1$}
\put(50,14){\kr}
\put(46,0){$\alpha_3$}
\end{picture}}
\hspace{-175 pt}
C=\left(\begin{smallmatrix}
                          2&-3&0\\-1&2&0\\0&0&2
                     \end{smallmatrix}\right),
\qquad
C^{-1}=\tfrac12\left(\begin{smallmatrix}
                          4&6&0\\2&4&0\\0&0&1
                     \end{smallmatrix}\right).
\end{gather*}

Hence $\alpha_2$ is the shorter of the simple roots. The relative
lengths of the simple roots are set as
$\langle\alpha_2,\alpha_2\rangle=\tfrac23$ and $\langle\alpha_1
,\alpha_1\rangle=\langle\alpha_3 ,\alpha_3\rangle=2$.

The bases of simple roots and fundamental weights are thus related by
\begin{gather*}
\begin{array}{llll}
\alpha_1=2\omega_1-3\omega_2,&\quad
\omega_1=2\alpha_1+3\alpha_2,&\quad
\check\alpha_1=\alpha_1,&\quad
\check\omega_1=\omega_1,
\\
\alpha_2=-\omega_1+2\omega_2,&\quad
\omega_2=\alpha_1+2\alpha_2,&\quad
\check\alpha_2=3\alpha_2,&\quad
\check\omega_2=3\omega_2,
\\
\alpha_3=2\omega_3;&\quad
\omega_3=\tfrac12\alpha_3;&\quad
\check\alpha_3=\alpha_3;&\quad
\check\omega_3=\omega_3.
\end{array}
\end{gather*}

Relative to the orthonormal basis, we have
\begin{gather*}
\begin{array}{llll}
\alpha_1=(\sqrt 2,0,0)=\sqrt 2e_1,&\quad
\omega_1=(\tfrac{1}{\sqrt 2},\tfrac{\sqrt 3}{\sqrt 2},0)=\tfrac{1}{\sqrt 2}e_1+\tfrac{\sqrt 3}{\sqrt 2}e_2,&\quad
\check\omega_1=(\tfrac{1}{\sqrt 2},\tfrac{\sqrt 3}{\sqrt 2},0),
\\
\alpha_2=(-\tfrac{1}{\sqrt 2},\tfrac{1}{\sqrt 6},0)=-\tfrac{1}{\sqrt 2}e_1+\tfrac{1}{\sqrt 6}e_2,&\quad
\omega_2=(0,\tfrac{\sqrt 2}{\sqrt 3},0)=\tfrac{\sqrt 2}{\sqrt 3}e_2,&\quad
\check\omega_2=(0,\sqrt6,0),
\\
\alpha_3=(0,0,\sqrt 2)=\sqrt 2e_3;&\quad
\omega_3=(0,0,\tfrac{1}{\sqrt 2})=\tfrac{1}{\sqrt 2}e_3;&\quad
\check\omega_3=(0,0,\tfrac{1}{\sqrt 2}).
\end{array}
\end{gather*}

The highest roots for the simple subgroups are given by the formulas
\begin{gather*}
\xi=2\alpha_1+3\alpha_2,
\qquad
\xi=\alpha_3.
\end{gather*}

The fundamental region $F$ is a cylinder (see Fig.~\ref{fig_g2a1}a) with a triangular base. Its~vertices~are
\begin{gather*}
F(G_2\times A_1)=\{0,\ \tfrac12 \cw_1,\ \tfrac13\cw_2,\ \cw_3,\ \tfrac12\cw_1+\cw_3,\ \tfrac13\cw_2+\cw_3\}.
\end{gather*}

The volume of the fundamental region is given by the formula
\begin{gather*}
|F(G_2\times A_1)|= \tfrac12\l[\tfrac 12\cw_1,\tfrac 13\cw_2],\cw_3\r=
\tfrac 12\cdot \tfrac{1}{2\sqrt 2}\cdot\tfrac{1}{\sqrt 3}\cdot\tfrac{1}{\sqrt 2}
\left|\begin{smallmatrix}
1&\sqrt3&0\\
0&\sqrt2&0\\
0&0&1\\
\end{smallmatrix}\right|=
\frac{\sqrt 6}{24},\quad
\text{where}\quad [\cdot,\cdot]\quad \text{denotes the vector product}.
\end{gather*}

The grid $F_{M,M'}\subset F$ is fixed by the independent choice of two positive integers $M$ and $M'$.
It consists of all the points
\begin{gather*}
F_{M,M'}(G_2\times A_1)=\left\{\tfrac{s_1}{M}\cw_1+\tfrac{s_2}{M}\cw_2+\tfrac{s'_1}{M'}\cw_3\mid\
2s_1+3s_2 \le M,\ s'_1 \le M';\ s_1,s_2,s_1'\in \Z^{\ge 0}\right\}.
\end{gather*}

The number of points in the grid $F_{M,M'}$ equals to
\begin{gather*}
|F_{M,M''}(G_2\times A_1)|=|F_{M}(G_2)|\cdot|F_{M'}(A_1)|=
\left(\left[\tfrac{M}{3}\right]+1+\sum_{i=0}^{\frac M3}\left[\tfrac{M-3i}{2}\right]\right)(M'+1),
\end{gather*}
where $[\ \cdot\ ]$ is the integer part of a number.

\begin{example}\label{ex_grid_g2a1}
Consider the case $M=4$, $M'=3$. There are 16 points of $F_{4,3}$ (see Fig.~\ref{fig_g2a1}b)
\begin{gather*}
|F_{4,3}(G_2\times A_1)|=
\left(
\left[\tfrac{4}{3}\right]+1+\sum_{i=0}^{\left[\frac{4}{3}\right]}\left[\tfrac{4-3i}{2}\right]
\right)
(3+1)=(1+1+(2+0))\cdot 4=16.
\end{gather*}
Explicitly, we have the following sets of integers $[s_1,s_2,s_1']$
and the corresponding grid points in the $\cw$-basis,
$(\tfrac{s_1}M,\tfrac{s_2}{M},\tfrac{s_1'}{M'})$:
\begin{gather*}
\begin{array}{llll}
[0,0,0]=(0,0,0),&\;
[0,0,1]=(0,0,\tfrac13),&\;
[0,0,2]=(0,0,\tfrac23),&\;
[0,0,3]=(0,0,1),
\\[1pt]
[0,1,0]=(0,\tfrac14,0),&\;
[0,1,1]=(0,\tfrac14,\tfrac13),&\;
[0,1,2]=(0,\tfrac14,\tfrac23),&\;
[0,1,3]=(0,\tfrac14,1),
\\[1pt]
[1,0,0]=(\tfrac14,0,0),&\;
[1,0,1]=(\tfrac14,0,\tfrac13),&\;
[1,0,2]=(\tfrac14,0,\tfrac23),&\;
[1,0,3]=(\tfrac14,0,1),
\\[1pt]
[2,0,0]=(\tfrac12,0,0),&\;
[2,0,1]=(\tfrac12,0,\tfrac13),&\;
[2,0,2]=(\tfrac12,0,\tfrac23),&\;
[2,0,3]=(\tfrac12,0,1).
\end{array}
\end{gather*}

Next, consider the case  $M=3$, $M'=2$.
$F_{3,2}(G_2\times A_1)$ consists of the following points:
\begin{gather*}
\begin{array}{lll}
[0,0,0]=(0,0,0),&\quad
[0,0,1]=(0,0,\tfrac12),&\quad
[0,0,2]=(0,0,1),
\\[1pt]
[1,0,0]=(\tfrac13,0,0),&\quad
[1,0,1]=(\tfrac13,0,\tfrac12),&\quad
[1,0,2]=(\tfrac13,0,1),
\\[1pt]
[0,1,0]=(0,\tfrac13,0),&\quad
[0,1,1]=(0,\tfrac13,\tfrac12),&\quad
[0,1,2]=(0,\tfrac13,1).
\end{array}
\end{gather*}
In this case, none of the nine points of $F_{3,2}(G_2\times A_1)$
coincides with a point of $F_{4,3}(G_2\times A_1)$.
This is due to the fact that lattice densities $M=3$ and $M'=2$
do not correspondingly divide the densities $M=4$ and $M'=3$.
\end{example}

The Weyl group orbit of the generic point $a\omega_1+b\omega_2+c\omega_3$, $\ a,b,c>0$,
consists of 24 points
\begin{gather*}
W_{(a,b,c)}(G_2\times A_1)=\{
\pm(a, b, c),\
\pm(-a, 3a+b, c),\
\pm(a+b, -b, c),\
\pm(2a+b, -(3a+b), c),
\\
\quad
\pm(-(a+b), 3a+2b, c),\
\pm(-(2a+b), 3a+2b, c),
\\
\qquad\qquad\qquad\qquad\qquad\qquad\;
\pm(a, b, -c),\
\pm(-a, 3a+b, -c),\
\pm(a+b, -b, -c),\
\pm(2a+b, -(3a+b), -c),
\\
\quad\quad\quad
\pm(-(a+b), 3a+2b, -c),\
\pm(-(2a+b), 3a+2b, -c)
\}.
\end{gather*}

Orbit sizes for arbitrary points are given by the relations
\begin{gather*}
\begin{array}{llll}
|W_{(a,b,c)}|=24,&\quad
|W_{(a,b,0)}|=12,&\quad
|W_{(a,0,c)}|=12,&\quad
|W_{(0,b,c)}|=12,
\\[1pt]
|W_{(a,0,0)}|=6,&\quad
|W_{(0,b,0)}|=6,&\quad
|W_{(0,0,c)}|=2,&\quad
|W_{(0,0,0)}|=1.
\end{array}
\end{gather*}

\bigskip

\subsection{The Lie algebra ${A_3}$}\label{ssec_A3}\

\begin{figure}[h]
\centerline{\includegraphics[scale=0.88]{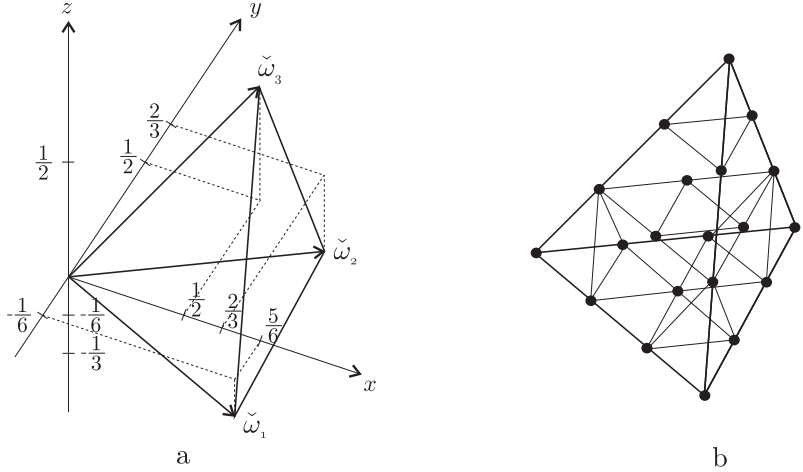}}
\caption{a)~the fundamental region $F$ of the Lie algebra $A_3$;
$x$, $y$ and $z$ indicate respectively orthogonal directions of the orthonormal basis $\{e_1$, $e_2$, $e_3\}$,
and the fundamental region of $A_3$ is the tetrahedron;
b) the grid $F_{3}(A_3)$.}\label{fig_a3}
\end{figure}

The Dynkin diagram and Cartan matrix with its inverse are the following,
\begin{gather*}
\parbox{.6\linewidth}
{\setlength{\unitlength}{1pt}
\def\kr{\circle{10}}
\def\cr{\circle*{10}}
\thicklines
\begin{picture}(20,30)
\put(8,24){$1$}
\put(10,14){\kr}
\put(6,0){$\alpha_1$}
\put(15,14){\line(1,0){10}}
\put(28,24){$1$}
\put(30,14){\kr}
\put(26,0){$\alpha_2$}
\put(35,14){\line(1,0){10}}
\put(48,24){$1$}
\put(50,14){\kr}
\put(46,0){$\alpha_3$}
\end{picture}}
\hspace{-175 pt}
C=\left(\begin{smallmatrix}
                          2&-1&0\\-1&2&-1\\0&-1&2
                     \end{smallmatrix}\right),
\qquad
C^{-1}=\tfrac14\left(\begin{smallmatrix}
                          3&2&1\\2&4&2\\1&2&3
                     \end{smallmatrix}\right).
\end{gather*}

Hence, all simple roots of the same length equal to~$\sqrt 2$.

The bases of simple roots and fundamental weights are thus related by
\begin{gather*}
\begin{array}{lll}
\alpha_1=2\omega_1-\omega_2,&\quad
\omega_1=\tfrac34\alpha_1+\tfrac12\alpha_2+\tfrac14\alpha_3,&\quad
\check\alpha_i=\alpha_i,\quad i\in \{1,2,3\}.
\\
\alpha_2=-\omega_1+2\omega_2-\omega_3,&\quad
\omega_2=\tfrac12\alpha_1+\alpha_2+\tfrac12\alpha_3,&\quad
\check\omega_i=\omega_i,\quad i\in\{1,2,3\}.
\\
\alpha_3=-\omega_2+2\omega_3;&\quad
\omega_3=\tfrac14\alpha_1+\tfrac12\alpha_2+\tfrac34\alpha_3;
\end{array}
\end{gather*}

Relative to the orthonormal basis, we have
\begin{gather*}
\begin{array}{lll}
\alpha_1=(1,-1,0)=e_1-e_2,&\quad
\omega_1=(\tfrac56,-\tfrac16,-\tfrac16)=\tfrac16(5e_1-e_2-e_3)=\check\omega_1,
\\
\alpha_2=(0,1,-1)=e_2-e_3,&\quad
\omega_2=(\tfrac23,\tfrac23,-\tfrac13)=\tfrac13(2e_1+2e_2-e_3)=\check\omega_2,
\\
\alpha_3=(\tfrac13,\tfrac13,\tfrac43)=\tfrac13(e_1+e_2+4e_3);&\quad
\omega_3=(\tfrac12,\tfrac12,\tfrac12)=\tfrac12(e_1+e_2+e_3)=\check\omega_3.
\end{array}
\end{gather*}

The highest root $\xi$ is given by the formula
\begin{gather*}
\xi=\alpha_1+\alpha_2+\alpha_3.
\end{gather*}

The fundamental region is a pyramid (see Fig.~\ref{fig_a3}a) with vertices
\begin{gather*}
F(A_3)=\{0,\ \cw_1,\ \cw_2,\ \cw_3\}.
\end{gather*}

The volume of the fundamental region is given by the formula
\begin{gather*}
|F(A_3)|=\tfrac 16 \l[\cw_1,\cw_2],\cw_3\r=\tfrac 16\cdot\tfrac 16\cdot\tfrac 13\cdot\tfrac 12\cdot
\left|\begin{smallmatrix}
5&-1&-1\\
2&2&-1\\
1&1&1\\
\end{smallmatrix}\right|=\tfrac1{12},
\end{gather*}
where $[\cdot,\cdot]$ denotes the vector product.

The grid $F_{M}\subset F$ is fixed by the choice of one positive integer $M$.
It consists of all the points
\begin{gather*}
F_{M}(A_3)=\left\{\tfrac{s_1}{M}\cw_1+\tfrac{s_2}{M}\cw_2+\tfrac{s_3}{M}\cw_3\mid\
s_1+s_2+s_3 \le M;\ s_1,s_2,s_3\in \Z^{\ge 0}\right\}.
\end{gather*}

The number of points in the grid $F_{M}$ equals to
\begin{gather*}
|F_{M}(A_3)|=\frac12\sum_{i=0}^{M}(M+1-i)(M+2-i).
\end{gather*}

\begin{example}\label{ex_grid_a3}
Consider the case $M=3$. There are 20 points of $F_{3}(A_3)$ (see Fig.~\ref{fig_a3}b).
Explicitly, we have the following sets of integers $[s_1,s_2,s_3]$
and the corresponding grid points in the $\cw$-basis
$(\tfrac{s_1}M,\tfrac{s_2}{M},\tfrac{s_3}{M})$:
\begin{gather*}
\begin{array}{llll}
[0,0,0]=(0,0,0),&\;
[0,0,3]=(0,0,1),&\;
[0,3,0]=(0,1,0),&\;
[3,0,0]=(1,0,0),
\\[1pt]
[2,0,1]=(\tfrac23,0,\tfrac13),&\;
[2,1,0]=(\tfrac23,\tfrac13,0),&\;
[1,0,2]=(\tfrac13,0,\tfrac23),&\;
[0,1,2]=(0,\tfrac13,\tfrac23),
\\[1pt]
[1,2,0]=(\tfrac13,\tfrac23,0),&\;
[0,2,1]=(0,\tfrac23,\tfrac13),&\;
[1,0,0]=(\tfrac13,0,0),&\;
[0,0,1]=(0,0,\tfrac13),
\\[1pt]
[0,1,0]=(0,\tfrac13,0),&\;
[1,1,0]=(\tfrac13,\tfrac13,0),&\;
[1,0,1]=(\tfrac13,0,\tfrac13),&\;
[0,1,1]=(0,\tfrac13,\tfrac13),
\\[1pt]
[1,1,1]=(\tfrac13,\tfrac13,\tfrac13),&\;
[2,0,0]=(\tfrac23,0,0),&\;
[0,0,2]=(0,0,\tfrac23),&\;
[0,2,0]=(0,\tfrac23,0).
\end{array}
\end{gather*}
\end{example}

The Weyl group orbit of the generic point $a\omega_1+b\omega_2+c\omega_3$, $\ a,b,c>0$,
consists of 24 points
\begin{gather*}
W_{(a,b,c)}(A_3)=\{
(a,\ b,\ c),
(-a,\ a+b,\ c),
(a+b,\ -b,\ b+c),
(a,\ b+c,\ -c),
(b,\ -(a+b),\ a+b+c),
\\ \hspace{-20pt}
(-a,\ a+b+c,\ -c),
(-(a+b),\ a,\ b+c),
(a+b,\ c,\ -(b+c)),
\\ \hspace{30pt}
(a+b+c,\ -(b+c),\ c),
(b,\ c,\ -(a+b+c)),
(b+c,\ -(a+b+c),\ a+b),
\\ \hspace{30pt}
(-b,\ -a,\ a+b+c),
(-(a+b),\ a+b+c,\ -(b+c)),
(-(a+b+c),\ a,\ b),
\\ \hspace{10pt}
(a+b+c,\ -c,\ -b),
(b+c,\ -c,\ -(a+b)),
(-b,\ b+c\ -(a+b+c)),
\\ \hspace{20pt}
(c,\ -(a+b+c),\ a),
(-(b+c),\ -a,\ (a+b)),
(-(a+b+c),\ a+b,\ -b),
\\ \hspace{48pt}
(c,\ -(b+c),\ -a),
(-(b+c),\ b,\ -(a+b)),
(-c,\ -(a+b),\ a),
(-c,\ -b,\ -a)
\}.
\end{gather*}

Orbit sizes for arbitrary points are given by the relations
\begin{gather*}
\begin{array}{llll}
|W_{(a,b,c)}|=24,&\quad
|W_{(a,b,0)}|=12,&\quad
|W_{(a,0,c)}|=12,&\quad
|W_{(0,b,c)}|=12,
\\[1pt]
|W_{(a,0,0)}|=4,&\quad
|W_{(0,b,0)}|=6,&\quad
|W_{(0,0,c)}|=4,&\quad
|W_{(0,0,0)}|=1.
\end{array}
\end{gather*}


\bigskip

\subsection{The Lie algebra ${B_3}$}\label{ssec_B3}\

\begin{figure}[h]
\centerline{\includegraphics[scale=0.88]{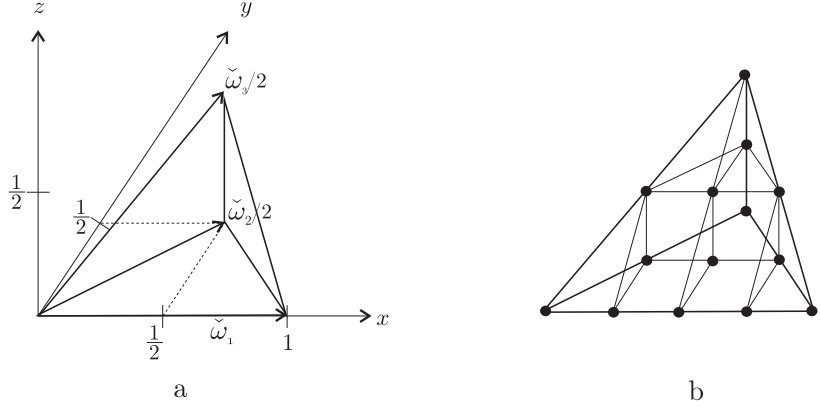}}
\caption{a)~the fundamental region $F$ of the Lie algebra $B_3$;
$x$, $y$ and $z$ indicate respectively orthogonal directions of the orthonormal basis $\{e_1$, $e_2$, $e_3\}$
and the face defined by vertices $\{0, \cw_1, \tfrac12\cw_2\}$ lies in the plane $z=0$;
b) the grid $F_{4}(B_3)$.}\label{fig_b3}
\end{figure}

The Dynkin diagram and Cartan matrix with its inverse are the following,
\begin{gather*}
\parbox{.6\linewidth}
{\setlength{\unitlength}{1pt}
\def\kr{\circle{10}}
\def\cr{\circle*{10}}
\thicklines
\begin{picture}(20,30)
\put(8,24){$1$}
\put(10,14){\kr}
\put(6,0){$\alpha_1$}
\put(15,14){\line(1,0){10}}
\put(28,24){$2$}
\put(30,14){\kr}
\put(26,0){$\alpha_2$}
\put(34,11){\line(1,0){13}}
\put(34,17){\line(1,0){13}}
\put(48,24){$2$}
\put(50,14){\cr}
\put(46,0){$\alpha_3$}
\end{picture}}
\hspace{-175 pt}
C=\left(\begin{smallmatrix}
                          2&-1&0\\-1&2&-2\\0&-1&2
                     \end{smallmatrix}\right),
\qquad
C^{-1}=\tfrac12\left(\begin{smallmatrix}
                          2&2&2\\2&4&4\\1&2&3
                     \end{smallmatrix}\right).
\end{gather*}

Hence $\alpha_1$ is the shorter of the simple roots.
The relative lengths of the simple roots are set as $\langle\alpha_1 ,\alpha_1\rangle=1$ and
$\langle\alpha_2 ,\alpha_2\rangle=\langle\alpha_3 ,\alpha_3\rangle=2$.

The bases of simple roots and fundamental weights are thus related by
\begin{gather*}
\begin{array}{llll}
\alpha_1=2\omega_1-\omega_2,&\quad
\omega_1=\alpha_1+\alpha_2+\alpha_3,&\quad
\check\alpha_1=\alpha_1,&\quad
\check\omega_1=\omega_1,
\\
\alpha_2=-\omega_1+2\omega_2-2\omega_3,&\quad
\omega_2=\alpha_1+2\alpha_2+2\alpha_3,&\quad
\check\alpha_2=\alpha_2,&\quad
\check\omega_2=\omega_2,
\\
\alpha_3=-\omega_2+2\omega_3;&\quad
\omega_3=\tfrac12\alpha_1+\alpha_2+\tfrac32\alpha_3;&\quad
\check\alpha_3=2\alpha_3;&\quad
\check\omega_3=2\omega_3.
\end{array}
\end{gather*}

Relative to the orthonormal basis, we have
\begin{gather*}
\begin{array}{llll}
\alpha_1=(1,-1,0)=e_1-e_2,&\quad
\omega_1=(1,0,0)=e_1,&\quad
\check\omega_1=(1,0,0)=e_1,
\\
\alpha_2=(0,1,-1)=e_2-e_3,&\quad
\omega_2=(1,1,0)=e_1+e_2,&\quad
\check\omega_2=(1,1,0)=e_1+e_2,
\\
\alpha_3=(0,0,1)=e_3;&\quad
\omega_3=(\tfrac12,\tfrac12,\tfrac12)=\tfrac12(e_1+e_2+e_3);&\quad
\check\omega_3=(1,1,1)=e_1+e_2+e_3.
\end{array}
\end{gather*}

The highest root $\xi$ is given by the formula
\begin{gather*}
\xi=\alpha_1+2\alpha_2+2\alpha_3.
\end{gather*}

The fundamental region is a pyramid (see Fig.~\ref{fig_a1a1a1}a) with vertices
\begin{gather*}
F(B_3)=\{0,\ \cw_1,\ \tfrac12 \cw_2,\ \tfrac12 \cw_3\}.
\end{gather*}

The volume of the fundamental region is given by the formula
\begin{gather*}
|F(B_3)|=\tfrac 16 \l[\cw_1,\tfrac12 \cw_2],\tfrac12 \cw_3\r=\tfrac16 \cdot \tfrac12\cdot \tfrac12
\left|\begin{smallmatrix}
1&0&0\\
1&1&0\\
1&1&1\\
\end{smallmatrix}\right|=\tfrac{1}{24},\quad
\text{where}\quad [\cdot,\cdot]\quad \text{denotes the vector product}.
\end{gather*}

The grid $F_{M}\subset F$ is fixed by the choice of an integer $M$.
It consists of all the points
\begin{gather*}
F_{M}(B_3)=\left\{\tfrac{s_1}{M}\cw_1+\tfrac{s_2}{M}\cw_2+\tfrac{s_3}{M}\cw_3\mid\
s_1+2s_2+2s_3\le M;\ s_1, s_2, s_3\in \Z^{\ge 0}\right\}.
\end{gather*}

The number of points in the grid $F_{M}$ equals to
\begin{gather*}
|F_{M}(B_3)|=\left(\left[\tfrac{M}{2}\right]+1\right)
\left(\left[\tfrac{M}{2}\right]\left[\tfrac{M+1}{2}\right]+M+1
-\tfrac{M+2}{2}\left[\tfrac{M}{2}\right]\right)
+\sum_{i=0}^{\left[\frac{M}{2}\right]}i^2.
\end{gather*}
where $[\ \cdot\ ]$ is the integer part of a number.

\begin{example}\label{ex_grid_b3}
Consider the case $M=4$, see Fig.~\ref{fig_b3}b.
There are 14 points of $F_4$
\begin{gather*}
|F_{4}(B_3)|=\left(\left[\tfrac{4}{2}\right]+1\right)
\left(\left[\tfrac{4}{2}\right]\left[\tfrac{4+1}{2}\right]+4+1
-\tfrac{4+2}{2}\left[\tfrac{4}{2}\right]\right)
+\sum_{i=0}^{\left[\frac{4}{2}\right]}i^2=
3(2\cdot2+4+1-3\cdot2)+0+1+4
=14.
\end{gather*}
Explicitly, we have the following sets of integers $[s_1,s_2,s_3]$
and the corresponding grid points in the $\cw$-basis,
$(\tfrac{s_1}M,\tfrac{s_2}{M},\tfrac{s_3}{M})$:
\begin{gather*}
\begin{array}{llll}
[0,0,0]=(0,0,0),&\;
[0,0,1]=(0,0,\tfrac14),&\;
[0,0,2]=(0,0,\tfrac12),&\;
[1,0,1]=(\tfrac14,0,\tfrac14),
\\[1pt]
[0,2,0]=(0,\tfrac12,0),&\;
[0,1,1]=(0,\tfrac14,\tfrac14),&\;
[0,1,2]=(0,\tfrac14, \tfrac12),&\;
[1,0,2]=(\tfrac14,0,\tfrac12),
\\[1pt]
[0,0,4]=(0,0,1),&\;
[0,0,3]=(0,0,\tfrac34),&\;
[1,1,0]=(\tfrac14,\tfrac14,0),&\;
[2,0,0]=(\tfrac12,0,0).
\\[1pt]
[0,1,0]=(0,\tfrac14,0),&\;
[1,0,0]=(\tfrac14,0,0),
\end{array}
\end{gather*}
\end{example}

The Weyl group orbit of the generic point $a\omega_1+b\omega_2+c\omega_3$, $\ a,b,c>0$,
consists of 48 points
\begin{gather*}
W_{(a,b,c)}(B_3)=\{
\pm (a,\ b,\ c),
\pm (-a,\ a+b,\ c),
\pm (a+b,\ -b,\ 2b+c),
\pm (a,\ b+c,\ -c),
\\ \qquad\qquad\qquad\qquad\qquad\;\;
\pm (b,\ -(a+b),\ 2a+2b+c),
\pm (-a,\ a+b+c,\ -c),
\pm (-(a+b),\ a,\ 2b+c),
\\ \qquad\quad\;
\pm (a+b,\ b+c,\ -(2b+c)),
\pm (a+b+c,\ -(b+c),\ 2b+c),
\\ \qquad\qquad\quad\quad\quad\;\;\;
\pm (b,\ a+b+c,\ -(2a+2b+c)),
\pm (b+c,\ -(a+b+c),\ 2a+2b+c),
\\ \hspace{44 pt}
\pm (-b,\ -a,\ 2a+2b+c),
\pm (-(a+b),\ a+2b+c,\ -(2b+c)),
\\ \hspace{20 pt}
\pm (a+2b+c,\ -(b+c),\ c),
\pm (-(a+b+c),\ a,\ 2b+c),
\\ \hspace{36 pt}
\pm (a+b+c,\ b,\ -(2b+c)),
\pm (a+2b+c,\ -(a+b+c),\ c),
\\  \hspace{76 pt}
\pm (b+c,\ a+b,\ -(2a+2b+c)),
\pm (-b,\ a+2b+c,\ -(2a+2b+c)),
\\ \hspace{64 pt}
\pm (b+c,\ -(a+2b+c),\ 2a+2b+c),
\pm (-(a+2b+c),\ a+b,\ c),
\\ \hspace{16 pt}
\pm (a+2b+c,\ -b,\ -c),
\pm (-(b+c),\ -a,\ 2a+2b+c),
\\ \hspace{-40 pt}
\pm (-(a+b+c),\ a+2b+c,\ -(2b+c))
\}.
\end{gather*}

Orbit sizes for arbitrary points are given by the relations
\begin{gather*}
\begin{array}{llll}
|W_{(a,b,c)}|=48,&\quad
|W_{(a,b,0)}|=24,&\quad
|W_{(a,0,c)}|=24,&\quad
|W_{(0,b,c)}|=24,
\\[1pt]
|W_{(a,0,0)}|=6,&\quad
|W_{(0,b,0)}|=12,&\quad
|W_{(0,0,c)}|=8,&\quad
|W_{(0,0,0)}|=1.
\end{array}
\end{gather*}


\bigskip

\subsection{The Lie algebra ${C_3}$}\label{ssec_C3}\

\begin{figure}[h]
\centerline{\includegraphics[scale=0.88]{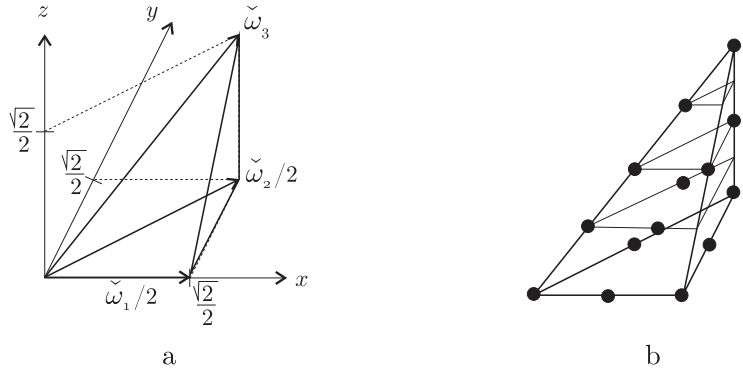}}
\caption{a)~the fundamental region $F$ of the Lie algebra $C_3$;
$x$, $y$ and $z$ indicate respectively orthogonal directions of the orthonormal basis $\{e_1$, $e_2$, $e_3\}$
and the face defined by vertices $\{0, \tfrac12\cw_1, \tfrac12\cw_2\}$ lies in the plane $z=0$;
b) the grid $F_{4}(C_3)$.}\label{fig_c3}
\end{figure}

The Dynkin diagram and Cartan matrix with its inverse are the following,
\begin{gather*}
\parbox{.6\linewidth}
{\setlength{\unitlength}{1pt}
\def\kr{\circle{10}}
\def\cr{\circle*{10}}
\thicklines
\begin{picture}(20,30)
\put(8,24){$2$}
\put(10,14){\cr}
\put(6,0){$\alpha_1$}
\put(15,14){\line(1,0){10}}
\put(28,24){$2$}
\put(30,14){\cr}
\put(26,0){$\alpha_2$}
\put(33,11){\line(1,0){13}}
\put(33,17){\line(1,0){13}}
\put(48,24){$1$}
\put(50,14){\kr}
\put(46,0){$\alpha_3$}
\end{picture}}
\hspace{-175 pt}
C=\left(\begin{smallmatrix}
                          2&-1&0\\-1&2&-1\\0&-2&2
                     \end{smallmatrix}\right),
\qquad
C^{-1}=\tfrac12\left(\begin{smallmatrix}
                          2&2&1\\2&4&2\\2&4&3
                     \end{smallmatrix}\right).
\end{gather*}

Hence $\alpha_3$ is the longer of the simple roots.
The relative lengths of the simple roots are set as
\mbox{$\langle\alpha_1 ,\alpha_1\rangle=\langle\alpha_2 ,\alpha_2\rangle=1$} and
\mbox{$\langle\alpha_3 ,\alpha_3\rangle=2$}.

The bases of simple roots and fundamental weights are thus related by
\begin{gather*}
\begin{array}{llll}
\alpha_1=2\omega_1-\omega_2,&\quad
\omega_1=\alpha_1+\alpha_2+\tfrac12\alpha_3,&\quad
\check\alpha_1=2\alpha_1,&\quad
\check\omega_1=2\omega_1,
\\
\alpha_2=-\omega_1+2\omega_2-\omega_3,&\quad
\omega_2=\alpha_1+2\alpha_2+\alpha_3,&\quad
\check\alpha_2=2\alpha_2,&\quad
\check\omega_2=2\omega_2,
\\
\alpha_3=-2\omega_1+2\omega_3;&\quad
\omega_3=\alpha_1+2\alpha_2+\tfrac32\alpha_3;&\quad
\check\alpha_3=\alpha_3;&\quad
\check\omega_3=\omega_3.
\end{array}
\end{gather*}

In the orthonormal basis these bases have the form
\begin{gather*}
\begin{array}{llll}
\alpha_1=(\tfrac1{\sqrt 2},-\tfrac1{\sqrt 2},0)=\tfrac1{\sqrt 2}(e_1-e_2),&\quad
\omega_1=(\tfrac1{\sqrt 2},0,0)=\tfrac1{\sqrt 2}e_1,&\quad
\check\omega_1=(\sqrt 2,0,0),
\\
\alpha_2=(0,\tfrac1{\sqrt 2},-\tfrac1{\sqrt 2})=\tfrac1{\sqrt 2}(e_2-e_3),&\quad
\omega_2=(\tfrac1{\sqrt 2},\tfrac1{\sqrt 2},0)=\tfrac1{\sqrt 2}(e_1+e_2),&\quad
\check\omega_2=(\sqrt 2,\sqrt 2,0),
\\
\alpha_3=(0,0,{\sqrt 2})={\sqrt 2}e_3;&\quad
\omega_3=(\tfrac1{\sqrt 2},\tfrac1{\sqrt 2},\tfrac1{\sqrt 2})=\tfrac1{\sqrt 2}(e_1+e_2+e_3);&\quad
\check\omega_3=(\tfrac1{\sqrt 2},\tfrac1{\sqrt 2},\tfrac1{\sqrt 2}).
\end{array}
\end{gather*}

The highest root $\xi$ is given by the formula
\begin{gather*}
\xi=2\alpha_1+2\alpha_2+\alpha_3.
\end{gather*}

The fundamental region is a pyramid (see Fig.~\ref{fig_a1a1a1}a) with vertices
\begin{gather*}
F(C_3)=\{0,\ \tfrac12 \cw_1,\ \tfrac12\cw_2,\ \cw_3\}.
\end{gather*}

The volume of the fundamental region is given by the formula
\begin{gather*}
|F(C_3)|=\tfrac 16 \l[\tfrac12\cw_1,\tfrac12\cw_2],\cw_3\r=
\tfrac16\cdot\tfrac12\cdot\tfrac12\cdot\sqrt 2\cdot\sqrt 2\cdot\tfrac1{\sqrt 2}
\left|\begin{smallmatrix}
1&0&0\\
1&1&0\\
1&1&1\\
\end{smallmatrix}\right|=\frac{\sqrt 2}{24},
\end{gather*}
where $[\cdot,\cdot]$ is the vector product.

The grid $F_{M}\subset F$ is fixed by the choice of the integer $M$.
It consists of all the points
\begin{gather*}
F_{M}(C_3)=\left\{\tfrac{s_1}{M}\cw_1+\tfrac{s_2}{M}\cw_2+\tfrac{s_3}{M}\cw_3\mid\
2s_1+2s_2+s_3\le M;\ s_1, s_2, s_3\in \Z^{\ge 0}\right\}.
\end{gather*}

The number of points in the grid $F_{M}$ equals to
\begin{gather*}
|F_{M}(C_3)|=\left(\left[\tfrac{M}{2}\right]+1\right)
\left(\left[\tfrac{M}{2}\right]\left[\tfrac{M+1}{2}\right]+M+1
-\tfrac{M+2}{2}\left[\tfrac{M}{2}\right]\right)
+\sum_{i=0}^{\left[\frac{M}{2}\right]}i^2.
\end{gather*}
where $[\ \cdot\ ]$ is the integer part of a number.

\begin{example}\label{ex_grid_c3}
Consider the case $M=4$. There are 14 points of $F_4(C_3)$, see Fig.~\ref{fig_c3}b.
\begin{gather*}
|F_{4}(C_3)|=\left(\left[\tfrac{4}{2}\right]+1\right)
\left(\left[\tfrac{4}{2}\right]\left[\tfrac{4+1}{2}\right]+4+1
-\tfrac{4+2}{2}\left[\tfrac{4}{2}\right]\right)
+\sum_{i=0}^{\left[\frac{4}{2}\right]}i^2=14.
\end{gather*}
Explicitly, we have the following sets of integers $[s_1,s_2,s_3]$
and the corresponding grid points in the $\cw$-basis,
$(\tfrac{s_1}M,\tfrac{s_2}{M},\tfrac{s_3}{M})$:
\begin{gather*}
\begin{array}{llll}
[0,0,0]=(0,0,0),&\;
[0,0,1]=(0,0,\tfrac14),&\;
[0,0,2]=(0,0,\tfrac12),&\;
[0,0,3]=(0,0,\tfrac34),
\\[1pt]
[0,1,0]=(0,\tfrac14,0),&\;
[0,1,1]=(0,\tfrac14,\tfrac14),&\;
[0,0,4]=(0,0,1),&\;
[2,0,0]=(\tfrac12,0,0),
\\[1pt]
[1,0,0]=(\tfrac14,0,0),&\;
[1,0,1]=(1,0,\tfrac13),&\;
[1,0,2]=(1,0,\tfrac23),&\;
[1,1,2]=(1,1,\tfrac23).
\\[1pt]
[1,1,0]=(1,1,0),&\;
[1,1,1]=(1,1,\tfrac13),&\;
\end{array}
\end{gather*}
\end{example}

The Weyl group orbit of the generic point $a\omega_1+b\omega_2+c\omega_3$, $\ a,b,c>0$,
consists of 48 points
\begin{gather*}
W_{(a,b,c)}(C_3)=\{
\pm (a,\ b,\ c),\
\pm (-a,\ a+b,\ c),\
\pm (a+b,\ -b,\ b+c),\
\pm (a,\ b+2c,\ -c),\
\\ \hspace{90 pt}
\pm (b,\ -(a+b),\ a+b+c),\
\pm (-a,\ a+b+2c,\ -c),\
\pm (-(a+b),\ a,\ b+c),\
\\ \hspace{30 pt}
\pm (a+b,\ b+2c,\ -(b+c)),\
\pm (a+b+2c,\ -(b+2c),\ b+c),\
\\ \hspace{66 pt}
\pm (b,\ a+b+2c,\ -(a+b+c)),\
\pm (b+2c,\ -(a+b+2c),\ a+b+c),\
\\ \hspace{28 pt}
\pm (-b,\ -a,\ a+b+c),\
\pm (-(a+b),\ a+2b+2c,\ -(b+c)),\
\\ \hspace{22 pt}
\pm (a+2b+2c,\ -(b+2c),\ c),\
\pm (-(a+b+2c),\ a,\ b+c),\
\\ \hspace{38 pt}
\pm (a+b+2c,\ b,\ -(b+c)),\
\pm (a+2b+2c,\ -(a+b+2c),\ c),\
\\ \hspace{60 pt}
\pm (b+2c,\ a+b,\ -(a+b+c)),\
\pm (-b,\ a+2b+2c,\ -(a+b+c)),\
\\ \hspace{62 pt}
\pm (b+2c,\ -(a+2b+2c),\ a+b+c),\
\pm (-(a+2b+2c),\ a+b,\ c),\
\\ \hspace{10 pt}
\pm (a+2b+2c,\ -b,\ -c),\
\pm (-(b+2c),\ -a,\ a+b+c),\
\\ \hspace{-50 pt}
\pm (-(a+b+2c),\ a+2b+2c,\ -(b+c))
\}.
\end{gather*}

Orbit sizes for arbitrary points are given by the relations
\begin{gather*}
\begin{array}{llll}
|W_{(a,b,c)}|=48,&\quad
|W_{(a,b,0)}|=24,&\quad
|W_{(a,0,c)}|=24,&\quad
|W_{(0,b,c)}|=24,
\\[1pt]
|W_{(a,0,0)}|=6,&\quad
|W_{(0,b,0)}|=12,&\quad
|W_{(0,0,c)}|=8,&\quad
|W_{(0,0,0)}|=1.
\end{array}
\end{gather*}

\section{Orbit functions}\label{sec_orbit-funcs}

In this section, we define what we mean by $C$-, $S$- and $E$-functions,
specified by a given point $\lambda\in\Z^n$.
We also show some of the properties inherent to those functions.
Namely, the following properties are of interest, their pairwise
orthogonality (using the appropriate scalar product, an integral
over the fundamental region), their discrete orthogonality (again,
using a properly defined scalar product, a sum over the discrete
grid), their product can be represented as a sum, and they are
eigenfunctions of the Laplace operator.

\subsection{Definitions, symmetries and general properties}\label{ssec_def_orb_funcs}\

We start with the $C$-functions. The $C$-function $C_{\lambda}(x)$, $\lambda\in P^+$ is defined as
\begin{gather}\label{def_c-function}
C_\lambda(x) := \sum_{\mu\in W_\lambda} e^{2\pi i \l\mu, x\r},
\qquad
x\in\R^n,
\end{gather}
where $W_\lambda$ is the Weyl group orbit generated from $\lambda$.

If in~(\ref{def_c-function}) we restrict ourselves to the orbit of the even subgroup $W_{e\lambda}$,
then we define \mbox{$E$-function} $E_{\lambda}(x)$, $\lambda\in P_e$
\begin{gather}\label{def_e-function}
E_\lambda(x) := \sum_{\mu\in W_{e\lambda}} e^{2\pi i \l\mu, x\r},
\qquad
x\in\R^n.
\end{gather}

The definition of an $S$-function $S_{\lambda}(x)$, $\lambda\in P^{++}$ is almost identical,
but the sign of each summand is determined by the number of
reflections $p(\mu)$ necessary to obtain $\mu$ from $\lambda$
\begin{gather}\label{def_s-function}
S_\lambda(x) := \sum_{\mu\in W_\lambda} (-1)^{p(\mu)}e^{2\pi i \l\mu , x\r}
\qquad
x\in\R^n.
\end{gather}
Of course the same $\mu$ can be obtained by different successions of reflections,
but all routes from $\lambda$ to $\mu$ will have a length of the same
parity, and thus the salient detail given by $p(\mu)$, in the
context of an $S$-function, is meaningful and unchanging.

For different families of orbit functions,
the $\lambda$ (represented in the $\omega$-basis) are taken
from different sets, namely
\begin{gather*}
\begin{array}{ll}
\lambda\in\{\Z^{\ge 0} \omega_1 + \Z^{\ge 0} \omega_2 + \Z^{\ge 0} \omega_3\},
&
\text{for}\; C\;\text{-functions};
\\
\lambda\in\{\Z^{\ge 0} \omega_1 + \Z^{\ge 0} \omega_2 + \Z^{\ge 0} \omega_3\}\cup
r_i\{\Z^{>0} \omega_1 + \Z^{>0} \omega_2 + \Z^{>0} \omega_3\},\; r_i\in W
&
\text{for}\; E\;\text{-functions};
\\
\lambda\in\{\Z^{>0} \omega_1 + \Z^{>0} \omega_2 + \Z^{>0} \omega_3\},
&
\text{for}\; S\;\text{-functions}.
\end{array}
\end{gather*}

In particular, this implies that, for $S$-functions,
the number of summands always equals to the size of the Weyl group.

In the case of $x\in F_M$ (the coordinates of $x$ are rational),
the $C$-, $S$- and $E$-functions are formed by roots of unity.
Therefore, there can only be a finite number of possible orbit functions
that can take distinct values on the points of $F_M$, these functions are orthogonal on the grid.
The number of pairwise orthogonal orbit functions on $F_M$ coincides with
the size of $F_M$, including the boundary in the case of $C$- and $E$-functions
and excluding the boundary in the case of , $S$-functions.

Note that in the 1-dimensional case, $C$-, $S$- and $E$-functions are respectively a cosine,
a sine and an exponential functions up to the constant.

All three families of orbit functions are based on semisimple Lie groups of finite order,
the number of variables coincides with the rank of the corresponding Lie algebra.

In general, $C$-, $S$- and $E$-functions are the finite sums of exponential functions,
therefore they are continuous and have continuous derivatives of all orders in $\R^n$.

It is easy to prove that $C$- and  $S$-functions are invariant with respect to
the action of both Weyl $W$ and affine Weyl $W^{aff}$ groups
(see e.g.~\cite{KlimykPatera2006, KlimykPatera2007-1})
and that $E$-functions are invariant with respect to
the action of $W_e$ and $W_e^{aff}$ groups (see e.g.~\cite{KashubaPatera2007}).
Therefore it is enough to consider them only on the fundamental domain
of their affine Weyl symmetry groups.

The $S$-functions are antisymmetric with respect to $n-1$-dimensional boundary of $F$.
Hence they are zero on the boundary of~$F$.
The $C$-functions are symmetric with respect to $n-1$-dimensional boundary of $F$.
Their normal derivative at the boundary is equal to zero (because the normal
derivative of a $C$-function is an $S$-function).
A number of other properties of orbit functions are presented
in~\cite{KlimykPatera2006, KlimykPatera2007-1, KP3}.

\subsection{Calculation of scalar products}\

Here we present the rules necessary for the calculation of the scalar products of the vectors given in the different bases.

Let vectors $u=(u_1,u_2,\dots,u_n)=u_1\w_1+u_2\w_2+\dots+u_n\w_n$
and $v= v_1\w_1+v_2\w_2+\dots+v_n\w_n$ are represented in the $\omega$-basis,
then their scalar product is calculated as the matrix product
\begin{gather*}
\l u , v\r = u \tilde C v^t=\sum_{i,j=1}^{n}u_i v_j\l\w_i,\w_j\r=
\sum_{i,j=1}^{n}u_i v_j\frac{\l\alpha_j,\alpha_j\r}2C^{-1}_{i,j},
\end{gather*}
here $\tilde C_{i,j}=\frac{\l\alpha_j,\alpha_j\r}2C^{-1}_{i,j}$ and $C$ is the Cartan matrix.

If vectors $x=(x_1,x_2,\dots,x_n)=x_1\alpha_1+x_2\alpha_2+\dots+x_n\alpha_n$
and $y= y_1\alpha_1+y_2\alpha_2+\dots+y_n\alpha_n$ are represented in the $\alpha$-basis,
then their scalar product is calculated as the matrix product
\begin{gather*}
\l x , y\r = x \hat C y^t=\sum_{i,j=1}^{n}x_i y_j\l\alpha_i,\alpha_j\r=
\sum_{i,j=1}^{n}x_i y_j\frac{\l\alpha_j,\alpha_j\r}2C_{i,j},
\end{gather*}
here $\hat C_{i,j}=\frac{\l\alpha_j,\alpha_j\r}2C_{i,j}$ and $C$ is the Cartan matrix.

Square lengths of the simple roots $\l\alpha_j,\alpha_j\r$ are always indicated
in the Dynkin diagrams of the semisimple Lie groups, see e.g.~\cite{Humphreys1972, KassMoodyPateraSlansky1990}.

It is also useful to remind that
\begin{gather*}
\l \alpha_i , \check\omega_j\r = \l \check \alpha_i , \omega_j\r=\l e_i , e_j\r=\delta_{i,j},
\quad \text{where}\;\delta_{i,j}\;\text{is the Kronecker delta}.
\end{gather*}

\subsection{$C$-, $S$-, and $E$-functions as eigenfunctions of the Laplace operator}\
Consider the functions $C_\lambda(x)$, $E_\lambda(x)$ and
$S_\lambda(x)$ and suppose that the continuous variable $x$ is given
relative to the orthogonal basis. In the case of Lie algebra $A_n$
we use orthogonal coordinates $x_1, x_2, \dots,x_{n+1}$ and
coordinates $x_1, x_2, \dots,x_{n}$ for $B_n$, $C_n$ and $D_n$ (the
orthogonal bases for these algebras are well known and can be found
e.g. in~\cite{KlimykPatera2006}).

The Laplace operator in orthogonal coordinates has the form
\begin{gather*}
\Delta = \frac{\partial^2}{\partial {x_1}}+
\frac{\partial^2}{\partial {x_2}}+\dots+ \frac{\partial^2}{\partial
{x_k}}, \quad \text{where}\; k=n\;(\text{or}\;k=n+1\; \text{for}\;
A_n) .
\end{gather*}

For the algebras $A_n$, $B_n$, $C_n$ and $D_n$, the Laplace operator gives the same eigenvalues on every exponential
function summand of an orbit function with eigenvalue~$-4\pi\langle \lambda,\lambda\rangle$.

Hence, the functions $C_\lambda(x)$, $E_\lambda(x)$ and $S_\lambda(x)$ are eigenfunctions
of the Laplace operator:
\begin{gather*}
\Delta
\left(\begin{array}{c}
C_\lambda(x)\\
E_\lambda(x)\\
S_\lambda(x)\\
\end{array}\right)
=-4\pi^2 \l\lambda ,\lambda\r
\left(\begin{array}{c}
C_\lambda(x)\\
E_\lambda(x)\\
S_\lambda(x)\\
\end{array}\right).
\end{gather*}

Now we consider the continuous variable $x$ given relative to the
$\w$-basis. Let $\Delta$ denote the Laplace operator, where the
differentiation $\partial_{x_i}$ is made with respect to the direction given by $\omega_i$.
\begin{gather*}
\Delta = \sum_{i,j=1}^n \frac{C_{ij}}{\l\alpha_i ,\alpha_i\r}
         \partial_{x_i}\partial_{x_j},
\text{where}\; C\;  \text{is the Cartan matrix}.
\end{gather*}
It is known in Lie theory that the matrix of
scalar products of the simple roots is positive definite,
moreover our definition makes matrix $\frac{C_{ij}}{\l\alpha_i ,\alpha_i\r}$ symmetric,
hence it can be diagonalized and the Laplace operator could be transformed
to the sum of second derivatives by an appropriate change of variables.

Thereby, using the results of Section~\ref{sec_algebras} we can write the explicit forms
of the Laplace operators given in the $\w$-basis for all semisimple Lie algebras of rank 3
\begin{gather*}
\Delta=
\begin{cases}
\partial_{x_1}^2+\partial_{x_2}^2+\partial_{x_3}^2,
\quad & \text{for}\quad A_1\times A_1\times A_1;
\\
\partial_{x_1}^2-\p_{x_1}\p_{x_2}+\partial_{x_2}^2+\partial_{x_3}^2,
\quad &\text{for}\quad A_2\times A_1;
\\
2\partial_{x_1}^2-2\partial_{x_1}\partial_{x_2}+\partial_{x_2}^2+\partial_{x_3}^2,
\quad &\text{for}\quad C_2\times A_1;
\\
3\partial_{x_1}^2-5\partial_{x_1}\partial_{x_2}+\partial_{x_2}^2+\partial_{x_3}^2,
\quad &\text{for}\quad G_2\times A_1;
\\
\partial_{x_1}^2-\partial_{x_1}\partial_{x_2}+\partial_{x_2}^2-\partial_{x_2}\partial_{x_3}+\partial_{x_3}^2,
\quad &\text{for}\quad A_3;
\\
\partial_{x_1}^2-\partial_{x_1}\partial_{x_2}+\partial_{x_2}^2-2\partial_{x_2}\partial_{x_3}+2\partial_{x_3}^2,
\quad &\text{for}\quad B_3;
\\
2\partial_{x_1}^2-2\partial_{x_1}\partial_{x_2}+2\partial_{x_2}^2-2\partial_{x_2}\partial_{x_3}+\partial_{x_3}^2,
\quad &\text{for}\quad C_3.
\\
\end{cases}
\end{gather*}

\subsection{Continuous orthogonality}\

For any two squared integrable functions $\phi(x)$ and $\psi(x)$
defined on the fundamental region $\widetilde{F}$, we define a continuous scalar product
\begin{gather}\label{def_cont_scalar_product}
\l\phi(x),\psi(x)\r:=\int_{{F}}\phi(x)\overline{\psi(x)}{\rm d}x.
\end{gather}
Here, integration is carried out with respect to the Euclidean measure, the bar means complex conjugation
and $x\in {F}$, where ${F}$ is the fundamental region of either $W$ or $W_e$.

Any pair of orbit functions from the same family is orthogonal on the corresponding fundamental region
with respect to the introduced scalar product~(\ref{def_cont_scalar_product}), namely
\begin{gather}\label{cont_orthog c funcs}
\l C_{\lambda}(x),C_{\lambda'}(x)\r=|W_\lambda|\cdot|F|\cdot\delta_{\lambda\lambda'},
\\\label{cont_orthog s funcs}
\l S_{\lambda}(x),S_{\lambda'}(x)\r=|W|\cdot|F|\cdot\delta_{\lambda\lambda'},
\\\label{cont_orthog e funcs}
\l E_{\lambda}(x),E_{\lambda'}(x)\r=|W_{e\ \lambda}|\cdot|F_e|\cdot\delta_{\lambda\lambda'},
\end{gather}
where $\delta_{\lambda\lambda'}$ is the Kronecker delta, $|W|$ is the size of Weyl group,
$|W_{\lambda}|$ and $|W_{e\ \lambda}|$ are the sizes of Weyl group orbits,
and $|F|$ and $|F_e|$ are volumes of fundamental regions.
All of the necessary information for each semisimple Lie algebra of rank 3
can be found in Section~\ref{sec_algebras}.
In particular, $|F_e|=2|F|$ and $|W_{e\ \lambda}|$ is defined by formula~(\ref{even_subgr_of_W}).

Proof of the relations~(\ref{cont_orthog c funcs},\ref{cont_orthog s funcs},\ref{cont_orthog e funcs})
follows from the orthogonality of the usual exponential functions
and from the fact that a given weight $\mu\in P$ belongs to precisely one orbit function.

Therefore each family of orbit functions forms an orthogonal basis
in the Hilbert space of squared integrable functions ${\mathcal L}^2(F)$.
Hence functions given on $F$ can be expanded in terms of linear combinations
of $C$-, $S$- or $E$-functions.

\subsection{Discrete orthogonality}\

Let us denote the discrete grid of the fundamental region as $F_M$ in the general case,
even though in some cases it is determined by more than one positive integer
\begin{gather*}
F_M=
\begin{cases}
F_M \quad & \text{for}\; C\text{-functions},
\\
F_{M}\setminus\partial F\quad & \text{for}\; S\text{-functions},
\\
F_{e\,M} \quad & \text{for}\; E\text{-functions}.
\\
\end{cases}
\end{gather*}
A discrete scalar product of two functions $\phi(x)$ and $\psi(x)$ given on ${F_M}$
(including $C$-, $S$- and $E$-functions) is dependent on this grid and defined by the bilinear form
\begin{gather}\label{def_discr_scalar_product}
\l\phi(x),\psi(x)\r_M= \sum_{i=1}^{|{F}_M|}\varepsilon(x_i)\phi(x_i)\overline{\psi(x_i)},
\quad x_i\in{F}_M.
\end{gather}
Here $\varepsilon(x_i)$ is the number of points conjugate to $x_i$ on the maximal torus of the Lie group.
The value of $\varepsilon(x_i)$ is given by the formula
\begin{gather*}
\varepsilon(x_i)=
\begin{cases}
|W_{x_i}| \quad & \text{for}\; C\text{-functions},
\\
|W|\quad & \text{for}\; S\text{-functions},
\\
|W_{e\,x_i}| \quad & \text{for}\; E\text{-functions}.
\\
\end{cases}
\end{gather*}

Again, as in the continuous case, the $C$-, $S$- and $E$-functions are pairwise orthogonal, i.e.
\begin{gather}\label{discr_ortog c}
\l C_{\lambda}(x),C_{\lambda'}(x)\r_M=
\sum_{\substack{i=\overline{1,N}}}|W_{x_i}|C_{\lambda}(x_i)\overline{C_{\lambda'}(x_i)}=
|W_\lambda|\cdot|A_M|\cdot\delta_{\lambda\lambda'},
\\ \label{discr_ortog s}
\l S_{\lambda}(x),S_{\lambda'}(x)\r_M=
|W|\sum_{\substack{i=\overline{1,N}}}S_{\lambda}(x_i)\overline{S_{\lambda'}(x_i)}=
|W|\cdot|A_M|\cdot\delta_{\lambda\lambda'},
\\ \label{discr_ortog e}
\l E_{\lambda}(x),E_{\lambda'}(x)\r_M=
\sum_{\substack{i=\overline{1,N}}}|W_{e\ x_i}|\phi_{\lambda}(x_i)\overline{\phi_{\lambda'}(x_i)}=
|W_{e\ \lambda}|\cdot|A_M|\cdot\delta_{\lambda\lambda'}.
\end{gather}
Here $|W_{x_i}|$ and $|W_{e\ x_i}|$ denote the sizes of the orbits of the Weyl group and its even subgroup.

$A_M$ is the $W$-invariant Abelian subgroup of the maximal torus ${\mathbb T}$
of the simple compact group corresponding to $W$
\begin{gather*}
A_M:=\{wx|x\in {F_M},\; w\in W\}
\qquad \text{and}\qquad
|A_M|=\sum_{i=1}^{|F_M|}|W_{x_i}|.
\end{gather*}

Proof of the orthogonality relations can be found
in~\cite{MoodyPatera2006}.

All necessary data for the computation of the coefficients
and discrete scalar products is given in Section~\ref{sec_algebras}.

\section{$C$-, $S$- and $E$-transforms}\label{sec_c-s-e-transforms}
For different fixed $m\in R^n$ the set of exponential functions
$\{e^{2\pi i\l m,x\r}$, $x\in R^n\}$
determines continuous and discrete Fourier transforms on $\R^n$.
In much the same way, the orbit functions
(which are a symmetrized version of exponential functions defined in Section~\ref{sec_orbit-funcs})
determine an analogue of the Fourier transform.

In this section, we introduce the essentials of the continuous and discrete $C$-, $S$- and $E$-transforms.
The discrete transform can be used for the continuous interpolation of values of a function $f(x)$
between its given values on a grid $F_M$.

\subsection{Continuous transforms}\label{ssec_cont_transforms}\

Each continuous function on the fundamental region with continuous derivatives
can be expanded as the sum of $C$-, $S$- or $E$-functions.
Let $f(x)$ be a function defined on $F$ (or $F_e$ for $E$-functions), then it may be written that
\begin{gather}\label{cont_trans_c}
f(x)=\sum_{\substack{\lambda\in P^+}}c_\lambda C_\lambda(x),
\qquad
c_\lambda=|W_{\lambda}|^{-1}|F|^{-1}\l f(x),C_{\lambda}(x)\r;
\\\label{cont_trans_s}
f(x)=\sum_{\substack{\lambda\in P^{++}}}c_\lambda S_\lambda(x),
\qquad
c_\lambda=|W|^{-1}|F|^{-1}\l f(x),S_{\lambda}(x)\r;
\\\label{cont_trans_e}
f(x)=\sum_{\substack{\lambda\in P_e}}c_\lambda E_\lambda(x),
\qquad
c_\lambda=|W_{e\ \lambda}|^{-1}|F_e|^{-1}\l f(x),E_{\lambda}(x)\r.
\end{gather}
Here $\l \cdot,\cdot\r$ denotes the continuous scalar product of~(\ref{def_cont_scalar_product}).
Direct and inverse $C$-, $S$- and $E$-transforms of the function $f(x)$ are
in~(\ref{cont_trans_c}), (\ref{cont_trans_s}) and (\ref{cont_trans_e})
respectively.

\subsection{Discrete transforms}\label{ssec_discr_transforms}\




Let $\Lambda_M\in P$ be the maximal set of points, such that for any two $\lambda,\lambda'\in\Lambda_M$
the condition of discrete orthogonality holds for any of the families of orbit functions
in~((\ref{discr_ortog c}), (\ref{discr_ortog s}) or (\ref{discr_ortog e})).

Then we have the following discrete transforms for the function $f(x)$:
\begin{gather}\label{discr_trans_c}
f(x) =\sum_{\lambda\in \Lambda_M} b_\lambda C_\lambda(x),
\quad
x\in F_M,
\qquad
b_\lambda=\frac{\l f,C_\lambda\r_M}{\l C_\lambda,C_\lambda\r_M};
\\ \label{discr_trans_s}
f(x) =\sum_{\lambda\in \Lambda_M} b_\lambda S_\lambda(x),
\quad
x\in F_M,
\qquad
b_\lambda=\frac{\l f,S_\lambda\r_M}{\l S_\lambda,S_\lambda\r_M};
\\\label{discr_trans_e}
f(x) =\sum_{\lambda\in \Lambda_M} b_\lambda E_\lambda(x),
\quad
x\in F_{e\ M},
\qquad
b_\lambda=\frac{\l f,E_\lambda\r_M}{\l E_\lambda,E_\lambda\r_M}.
\end{gather}
Here $\l\cdot,\cdot\r_M$ denotes the discrete scalar product given of~(\ref{def_discr_scalar_product}).

\subsection{Continuous extensions}\label{ssec_cont_ext}\

Once the coefficients $b_\lambda$ of the expansions~(\ref{discr_trans_c})  (\ref{discr_trans_s}), (\ref{discr_trans_e})
are calculated, discrete variables  $x_i$ in ${F_M}$ may be replaced by continuous variables $x$ in ${F}$
\begin{gather*}
f_{cont}(x) :=\sum_{\lambda\in \Lambda_M} b_\lambda C_\lambda(x),
\qquad
x\in F;
\\
f_{cont}(x) :=\sum_{\lambda\in \Lambda_M} b_\lambda S_\lambda(x),
\qquad
x\in F;
\\
f_{cont}(x) :=\sum_{\lambda\in \Lambda_M} b_\lambda E_\lambda(x),
\qquad
x\in F_{e}.
\end{gather*}
The function $f_{cont}(x)$ smoothly interpolates the values of $f(x_i)$, $i=1,2,\dots,|F_M|$.
At the points $x_i$, we have the equality $f_{cont}(x_i)=f(x_i)$.


\begin{remark}
If we calculate more than one discrete transform on the same grid ${F_M}$ and
use the same set of orbit functions $\phi_\lambda(x)$, $\lambda\in\Lambda_M$,
then it is reasonable to pre-compute and save the matrix
\begin{gather*}
B =
\begin{pmatrix}
\frac{\phi_{\lambda_1}(x_1)}{\l \phi_{\lambda_1}, \phi_{\lambda_1}\r_M} & \hdots & \frac{\phi_{\lambda_1}(x_{|F_M|})}{\l \phi_{\lambda_1}, \phi_{\lambda_1}\r_M}\\
\vdots & \ddots & \vdots \\
\frac{\phi_{\lambda_n}(x_1)}{\l \phi_{\lambda_n}, \phi_{\lambda_n}\r_M} & \hdots & \frac{\phi_{\lambda_n}(x_N)}{\l \phi_{\lambda_n}, \phi_{\lambda_n}\r_M}
\end{pmatrix}.
\end{gather*}
This would save valuable computation time, especially for large $M$,
since the coefficients $b_\lambda$ are easily calculated as a matrix product
$\overrightarrow{b}=B\cdot \overrightarrow{f}^t$.

Moreover, the matrix $B$ does not depend on the function that is to be expanded into series,
therefore it need be calculated only once for each $M$ and can be repeatedly used.
\end{remark}

\section{Concluding remarks}\label{sec_conclusions}

\begin{itemize}
\item
Each of the transforms described here is based on a compact semisimple Lie group of rank 3.
All seven types of such Lie groups were considered.
Our goal was to provide the tools for the expansion of functions of 3 variables given on a bounded region $D$
of an Euclidean space $\R^3$.  The variables can be either continuous or discrete (lattice grid points).
The symmetry of the lattice is the Weyl group of the Lie group.
The bounded region $D$ has to be scaled to fit into the fundamental region $F$
of the corresponding Lie group. In case of functions given on a lattice grid,
the scaling has to be accompanied with the matching density of the grid points in $F$.
Fortunately, the formalism admits choosing any density one may need.
The scaling resulting in the inclusion $D\subset F$ is not unique.
Various options may be considered for specific functions.
\smallskip

\begin{figure}[h]
\centerline{\includegraphics[scale=0.45]{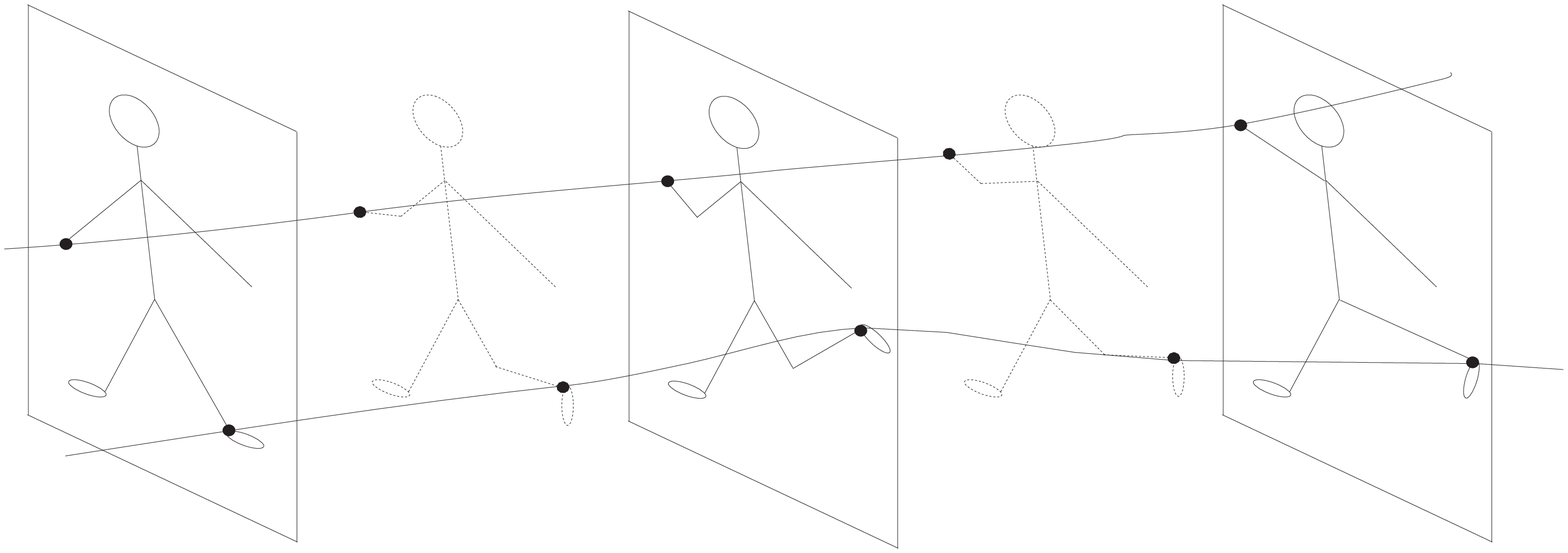}}
\caption{One of the possible applications of orbit functions is the construction of unknown transitional data
(smooth interpolation).
Fig.~\ref{interpolation} shows how additional frames could be added to a film.
In much the same way, the continuous deformation of the picture can be proceed and a
3-dimensional image can be created from corresponding 2-dimensional layers or cuts.
}\label{interpolation}
\end{figure}

\item
The uncommon special functions of our transforms are defined for compact semisimple
Lie groups of any type and rank \cite{patera2005}.
Their continuous and discrete orthogonality in $F$ is assured \cite{MoodyPatera2006}.
Unlike the translation symmetry required in traditional Fourier expansions,
the symmetry group of $C$-, $S$-, and $E$-functions is the appropriate affine Weyl group,
which contains the translations in $\R^3$ as a subgroup.
\smallskip

\item
The uniformity of our approach, as to the type of the rank 3 Lie group,
is illustrated here by considering the seven cases in parallel.
The price paid for uniformity is the exploitation of non-orthogonal bases,
$\alpha$, $\omega$-bases and their duals whenever necessary.
The majority of practically useful digital data usually given on
cubic/square lattices with the simplest symmetry group.
Only more costly experimental installations may use denser lattice arrangements of data collectors.

It should be pointed out that, at least for one type of transform, it is possible to avoid paying the price i.e. of having to work with orthonormal bases. The Weyl group of $SU(n)$
is isomorphic to the permutation group $S_n$ of $n$ elements.
Recently introduced transforms \cite{KlimykPatera2007-2,KlimykPatera2007-3,KP2008,KP6},
based on $S_n$ and on its alternating subgroup, exploit orthonormal bases in $\R^n$,
although even there, the corresponding fundamental regions do not have orthogonal adjacent faces, in general.

\item
There is an additional freedom of choice whenever the underlying Lie group is not simple.
Suppose that group is a product of two simple Lie groups,
$G^1\times G^2$. The fundamental region is then the Cartesian product of
$F(G^1)$ and $F(G^2)$.

For the expansion of class functions on the product group,
we can combine $C$-functions on one with $S$-functions on the other.
Similarly, we can combine $C$- or $S$-functions with $E$-functions,
enlarging appropriately the fundamental region of the $E$-functions.

In much the same way, discretization can proceed differently on $F(G^1)$ then on $F(G^2)$.
The corresponding integers $M_1$, $M_2$ that fix it can be as different as one desires.
Thus the density of grid points in $F(G^1)\times F(G^2)$ may be very different on the two orthogonal components.
\smallskip

\item
A number of other properties of the orbit functions may prove to be useful
(see  \cite{KlimykPatera2006,KlimykPatera2007-1,KP3} and references therein).
Let us point out that each of the three types of functions split
into mutually exclusive congruence classes. For a given semisimple Lie group,
the number of congruence classes equals the order of the center of the Lie group.
\smallskip

\item
The possibility to introduce the $C$-, $S$-, and $E$-functions
by summation over a finite noncrystallographic Coxeter groups instead of the Weyl group of a Lie group appears to be rather interesting.
In 3D there is just one such group $H_3$, the icosahedral group of order 120.
Most of the properties carry over to this case in a simple straightforward way.
The exception is the orthogonality, continuous or discrete.
There is an analog of the fundamental region, but no lattice.
Its role, perhaps, should be played by some quasicrystal?

\end{itemize}

\appendix
\section{Example of $C$-transform on $A_1\times A_1\times A_1$}\
As an example, we chose to interpolate a discretization of a known function,
namely the Gaussian function shown in equation~(\ref{eq_gauss})
\begin{gather}\label{eq_gauss}
g(x) = e^{-(x-p)^2} = e^{-(x-p)\cdot(x-p)},\quad x\in \R^3,
\end{gather}
where $p\in F(A_1\times A_1\times A_1)$ is a fixed point inside the fundamental region.

\begin{figure}[h]
 \centering
 \includegraphics[scale=0.95]{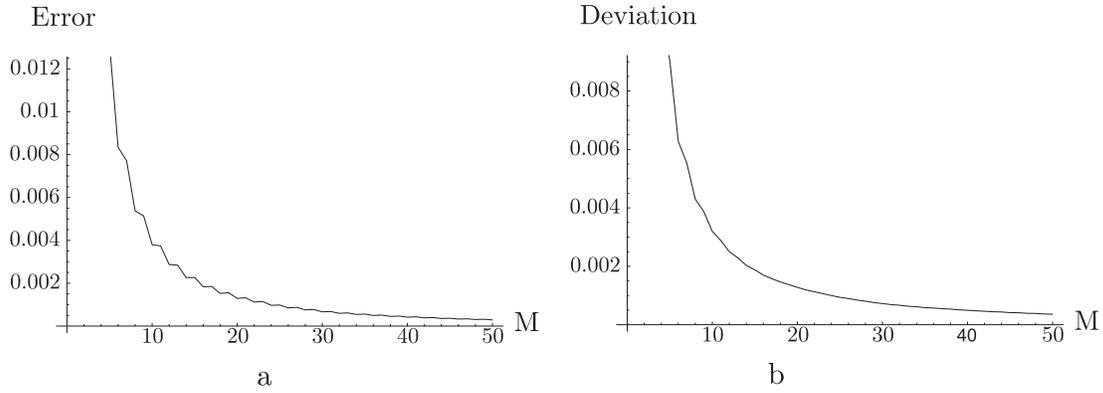}
 \caption{a)~the error~(\ref{eq_error}) of the $C$-transform~(\ref{eq_c-transform}) as function of $M$;
 b)~the standard deviation of the difference between the Gaussian function~(\ref{eq_gauss})
 and its interpolation~(\ref{eq_c-transform}) as function of $M$.}
 \label{figError}
\end{figure}

The first test to be undertaken is to sample the function in the points of the grid $F_M$
for several values of $M$.
The continuous extension of $g(x)$, calculated from points of $F_M$, is
\begin{gather}\label{eq_c-transform}
T(x) = \sum_{\lambda\in P} b_\lambda C_\lambda(x),
\quad \text{with} \quad
b_\lambda = \frac{\l g, C_\lambda\r}{\l C_\lambda, C_\lambda\r}.
\end{gather}
For different values of $M$, Fig.~\ref{figError}a represents the error, as defined by the integral
\begin{gather}\label{eq_error}
\int_{F} \left| T(x)-g(x)\right|d F.
\end{gather}
The integral (24) was calculated using a simple Monte Carlo method with 10000 randomly chosen points.
We made sure that the granularity of the randomly generated points was much
higher than that of the grid $F_M$.

Fig.~\ref{figError}b shows the standard deviation of the sample set.
It decreases in much the same way the error did in function of $M$.

\begin{figure}[h]
 \centering
 \includegraphics[scale=1]{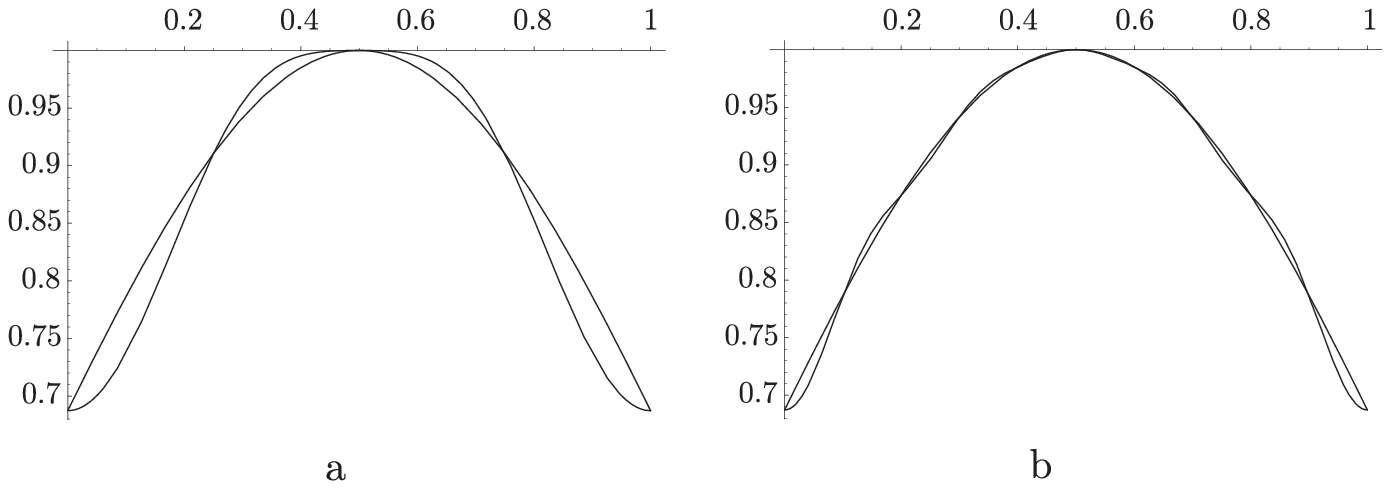}
 \caption{Comparison of $T\left( L(k)\right)$ (\ref{eq_c-transform}) with $g\left( L(k)\right)$
 for $k\in [0,1]$ (\ref{eq_parametrization}), with $M=4$ (a) and $M=10$ (b).}
 \label{cut1d}
\end{figure}

As a means to show what takes place visually, we give a plot of $T$ as
compared to $g$ on the parametric line given by the transformation
\begin{gather}\label{eq_parametrization}
L \colon \R \rightarrow \R^3,\quad
k \shortmid\!\xrightarrow{L}(k,k,k).
\end{gather}
Fig.~\ref{cut1d}a and Fig.~\ref{cut1d}b show the comparison between
$T\left( L(k)\right)$ and $g\left( L(k)\right)$ for $k\in [0,1]$,
with $M=4$ and $M=10$, respectively.
As seen when comparing Figs.~\ref{cut1d}a and~\ref{cut1d}b,
the degree of oscillation increases as $M$ increases, but the total error
decreases. This is exactly as one would expect.

Consider two-dimensional cuts on the parametric surface given by
\begin{gather}\label{eq_parametrization2}
S \colon \R^2 \rightarrow \R^3,\quad
(k,l) \shortmid\!\xrightarrow{L}(k+l,k+k,k+l)\quad
\text{with}\; k\in [0,\tfrac{1}{2}]\quad \text{and}\; l\in [0,\tfrac{1}{2}].
\end{gather}

\begin{figure}[h]
 \centering
 \includegraphics[scale=0.75]{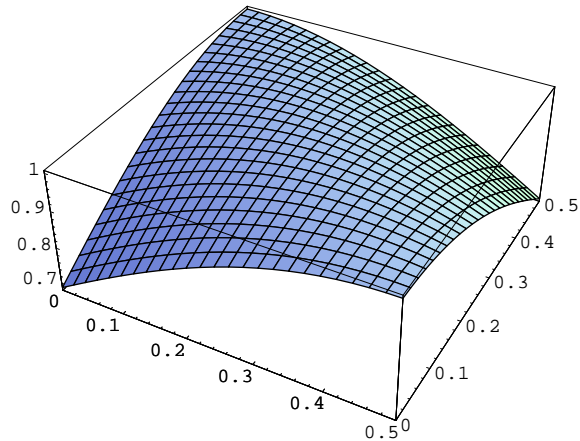}
 \caption{Two dimensional cut~(\ref{eq_parametrization2}) of the Gaussian function  $g\left( S(k,l)\right)$.}
 \label{gauss3d}
\end{figure}

\begin{figure}[h]
 \centering
 \includegraphics[scale=1]{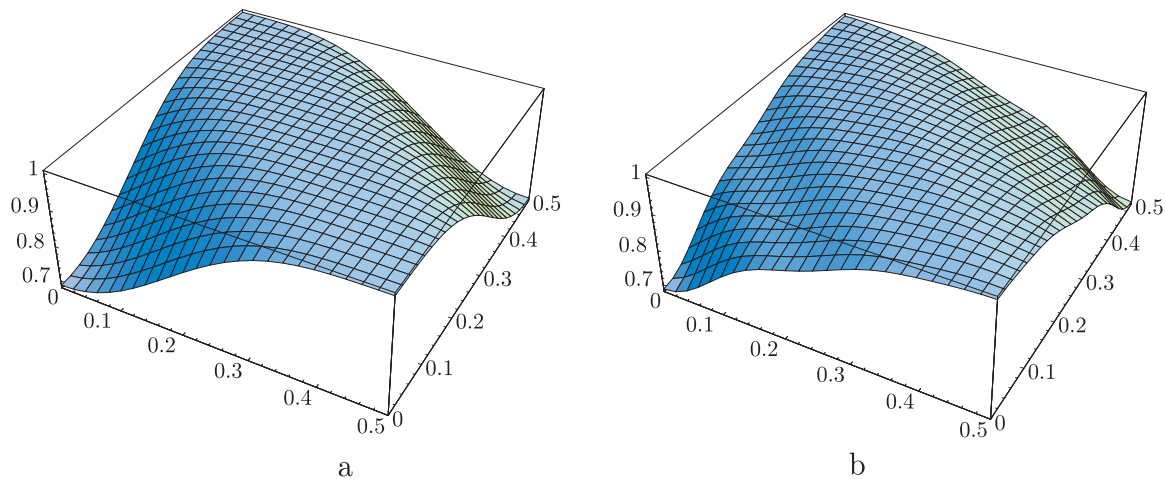}
 \caption{Two dimensional cut~(\ref{eq_parametrization2}) of the interpolation $T\left( S(k,l)\right)$
 for $M=4$~(a) and $M=10$~(b).}
 \label{surface}
\end{figure}

A two dimensional cut from the
graphs of $g$ is presented in Fig.~\ref{gauss3d}.
Figs.~\ref{surface}a and~\ref{surface}b are two dimensional cuts from the
graphs of $T$ with $M=4$ and $M=10$, respectively.

In the case of $A1\times A1\times A1$ (as well as in the other cases, albeit less trivially), the set of pairwise
orthogonal functions is not unique, i.e. one can scale all $\lambda\in P$ by integer multiplier  $s\cdot(a,b,c)$
with $(a,b,c)\in\Z^3, s\in \Z$.
This leads to a shifted system of orbit functions, that take the exact same values on the discrete grid $F_M$,
but do in fact differ in their continuous behavior.
Such a shifted system of functions can be understood as higher harmonics of the original functions on the grid $F_M$.

\begin{figure}[h]
 \centering
 \includegraphics[scale=1]{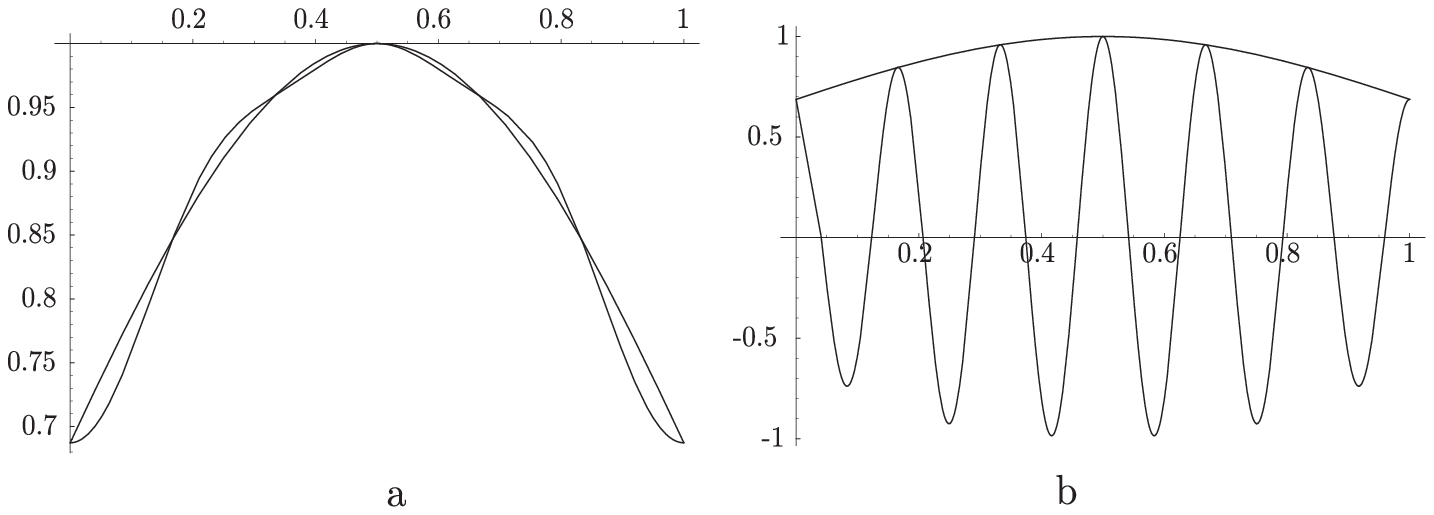}
 \caption{Comparison between  $g\left( L(k)\right)$
and
 $T\left( L(k)\right)$~(\ref{eq_c-transform})
 computed using the fundamental set of $C$-functions (a)
or computed using the shifted by a factor $(M,0,0)$ set of $C$-functions (b)
 for $k\in [0,1]$ (\ref{eq_parametrization}), with $M=6$.}
 \label{harmonics}
\end{figure}

Examples with original and shifted systems of orbit functions follow.
Fig.~\ref{harmonics}a shows $g$ and $T$ with $M=6$,
for $T$ computed using the fundamental set of $C$-functions
and Fig.~\ref{harmonics}b shows the same graph, but for a $T$ that was computed using the
set of $C$-functions shifted by a factor of $(M,0,0)$.

The result is enlightening.
The case using the shifted set of orbit functions 
exhibits a degree of error far above what could be tolerated,
due to the fact that the higher harmonics of the $C$-functions, as expected, oscillate more.
This derogatory result should not, however, dismiss the use of such higher harmonics.
The Gaussian function extrapolated here is very smooth, and should thus be extrapolated with
a sum of smoother functions.
But in real life applications, the data could be quite chaotic, thus the use
of higher harmonics could be useful.

It is also possible not to shift the entire set, but only a specific subset,
and at that, not all should be shifted by the same factor.
Careful consideration must be taken, because the functions do not obey the simplistic rule
that a shifted function is equal to its counterpart.
In fact, the set of adjacent functions is paired according to a reflective symmetry,
but not according to translation symmetry.

\subsection*{Acknowledgements}
Work supported in part by the Natural Sciences and Engineering Research Council of Canada,
the MIND Research Institute, by MITACS, and by Lockheed Martin Canada.
We are grateful for the hospitality extended to us at the Centre de recherches math\'ematiques,
Universit\'e de Montr\'eal (M.N.) and at the Aspen Center for Physics (J.P.) where most of this work was done.

\end{document}